\pdfoutput=1

\documentclass[11pt]{article}

\usepackage[]{acl} 

\usepackage{times}
\usepackage{latexsym}
\usepackage{amsmath}
\usepackage{hyperref}

\usepackage{placeins}
\usepackage[T1]{fontenc}

\usepackage[utf8]{inputenc}

\usepackage{microtype}

\usepackage{inconsolata}

\usepackage{booktabs}
\usepackage{adjustbox}
\usepackage{url}
\usepackage{soul,xcolor}
\usepackage{graphicx} 
\usepackage{amsmath}
\usepackage{fontawesome5}
\usepackage{multicol,multirow}
\usepackage{array}
\usepackage[capitalize]{cleveref}
\usepackage{enumitem}
\usepackage{tikz}
\usepackage{changepage}
\usepackage{amssymb}
\usepackage{relsize}
\usepackage{enumitem}
\usepackage{adjustbox}
\crefformat{section}{\S#2#1#3} 
\crefformat{subsection}{\S#2#1#3}
\crefformat{subsubsection}{\S#2#1#3}
\usepackage{pifont}
\usepackage{mdframed}

\let\realcite\cite
\renewcommand{\cite}[1]{\ifx.#1.\hl{[?]}\else\realcite{#1}\fi}
\newcommand{\cmark}{\ding{51}}%
\newcommand{\xmark}{\ding{55}}%
\newcommand{\dataset}{\texttt{MusicBench}}
\newcommand{\datasetFMA}{\texttt{FMACaps}}
\newcommand{\model}{\texttt{Mustango}}
\newcommand{\modelemoji}{\texttt{Mustango}}
\newcommand{\tabitem}{~~\llap{\textbullet}~~}

\newcommand{\modelfig}{\raisebox{-\ht\strutbox}{\includegraphics[width=0.06\textwidth]{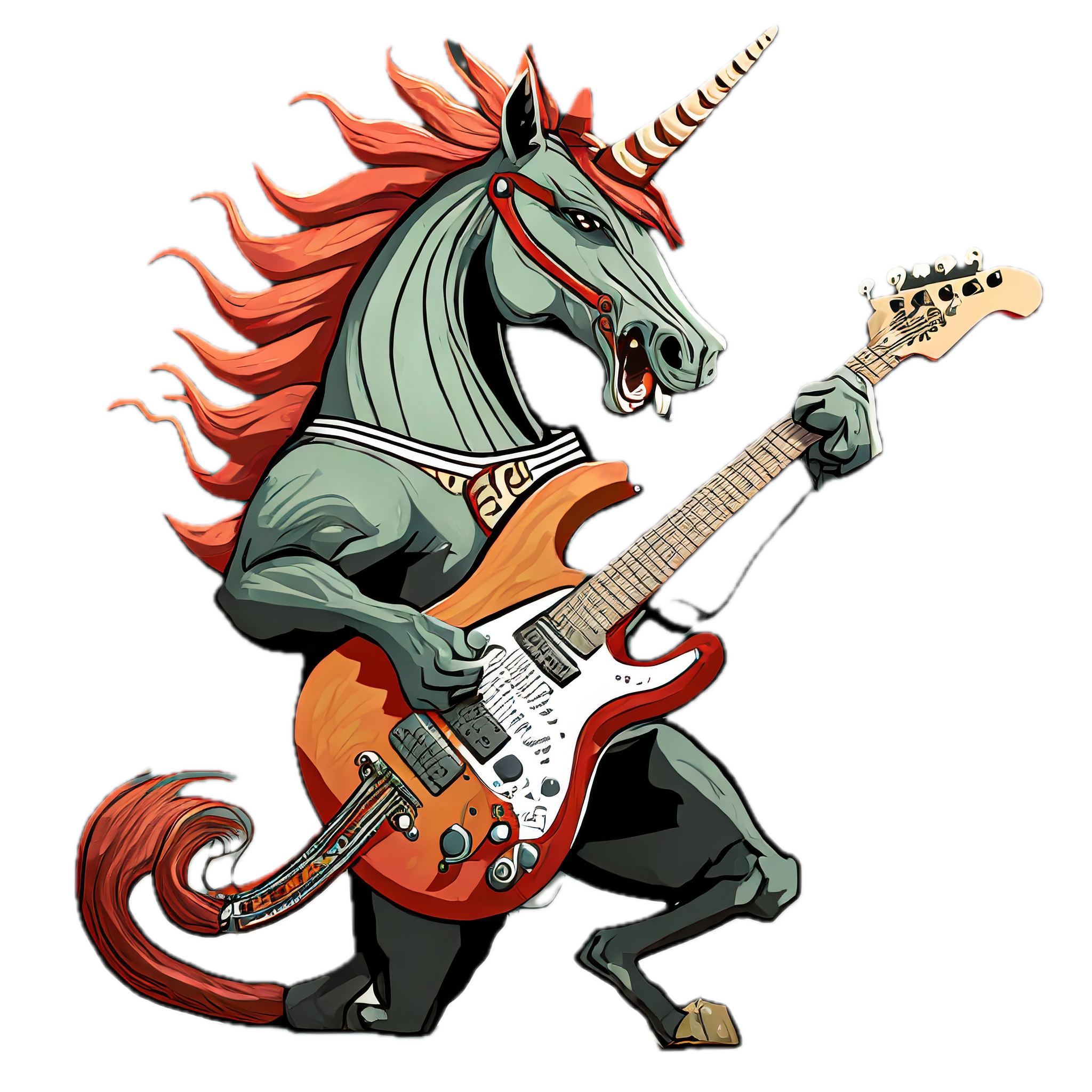}}
  \parbox{0.08\textwidth}{
   \model{}
  }}

\newcommand{\greencheck}{{\color{green}\cmark}}
\newcommand{\redcross}{{\color{red}\xmark}}

\newcommand\rurl[1]{%
  \href{https://#1}{\nolinkurl{#1}}%
}

\title{\hphantom{Mu}\model{}: Toward Controllable Text-to-Music Generation}

\author{Jan Melechovsky$^1$\thanks{\hspace{0.2cm} Co-first authors. Both authors contributed equally.} ,
  Zixun Guo$^2$\footnotemark[1] ,
  Deepanway Ghosal$^1$, \\
  \textbf{Navonil Majumder$^1$,
  Dorien Herremans$^1$\thanks{\hspace{0.2cm} Both authors contributed equally and led this project.},
  Soujanya Poria$^1$\footnotemark[2]}
  \\
  $^1$ Singapore University of Technology and Design, Singapore\\
  $^2$ Queen Mary University of London, UK
  }
  
\begin{document}
\maketitle
\begin{tikzpicture}[remember picture,overlay,shift={(current page.north west)}]
\node[anchor=north west,xshift=3.6cm,yshift=-2.2cm]{\scalebox{1}[1]{\includegraphics[width=1cm]{Sections/Figure/mustango2.png}}};
\end{tikzpicture}
\begin{minipage}[t]{2\linewidth}
\vspace{-1.75cm}
  \centering
  \faGithub: \url{https://github.com/AMAAI-Lab/mustango} \\
  \faGlobe : \url{https://huggingface.co/spaces/declare-lab/mustango}
\vspace{0.5cm}
\end{minipage}
\begin{abstract}


The quality of the text-to-music models has reached new heights due to recent advancements in diffusion models. The controllability of various musical aspects, however, has barely been explored.
In this paper, we propose \model{}: a music-domain-knowledge-inspired text-to-music system based on diffusion.  \model{} aims to control the generated music, not only with general text captions, but with more rich captions that can include specific instructions related to chords, beats, tempo, and key.
At the core of \model{} is MuNet, a Music-Domain-Knowledge-Informed UNet guidance module that steers the generated music to include the music-specific conditions, which we predict from the text prompt, as well as the general text embedding, during the reverse diffusion process.
To overcome the limited availability of open datasets of music with text captions, we propose a novel data augmentation method that includes altering the harmonic, rhythmic, and dynamic aspects of music audio and using state-of-the-art Music Information Retrieval methods to extract the music features which will then be appended to the existing descriptions in text format.
We release the resulting \dataset{} dataset which contains over 52K instances and includes music-theory-based descriptions in the caption text. 
Through extensive experiments, we show that the quality of the music generated by \model{} is state-of-the-art, and the controllability through music-specific text prompts greatly outperforms other models such as \texttt{MusicGen} and \texttt{AudioLDM2}.



\end{abstract}

\section{Introduction}
\label{sec:intro}


Recently, diffusion models~\cite{ho_diffusion} have shown prowess in image~\cite{dalle2}, audio~\cite{liu2023audioldm,liu2023audioldm2,ghosal2023tango, borsos2023audiolm} and music~\cite{huang2023noise2music, mousai} generation tasks. Generating music directly from a diffusion model poses some unique challenges. 
First, music adheres to specific rules related to, for instance, tempo, key, and chord progressions. Evaluating whether or not the generated music follows these conditions remains challenging. For instance, 
\texttt{MusicLM}~\cite{agostinelli2023musiclm}, a text-to-music model, ensures that the generated music matches the text prompts in terms of instrumentation and music vibe. However, the musicality of the generated music (e.g., musically meaningful harmonies and steady tempo) remains only partially addressed. Secondly, the availability of paired music and textual description datasets is limited~\cite{agostinelli2023musiclm, huang2023noise2music}. Although the textual descriptions in the existing datasets include details like instrumentation or vibe, more representational descriptions that capture the structural, melodic, and harmonic aspects of music are missing. We thus argue that including this information during generation may improve the current text-to-music models in terms of musicality and controllability (e.g., following metrical structure and chord progressions). More information on related work can be found in \cref{sec:relw}.
Beyond existing text-to-music systems' capability (e.g., setting correct instrumentation), our proposed \model{} model enables musicians, producers, and sound designers to create music clips with specific text-specified conditions like following a chord progression, setting tempo, and key selection.


In this paper, we release the \dataset{} dataset which is derived from the \texttt{MusicCaps}~\cite{agostinelli2023musiclm} dataset and propose \modelemoji{} to address these challenges. To create the \dataset{} dataset, we use two augmentation methods: \emph{description enrichment} and \emph{music diversification}. The aim of \emph{description enrichment} is to augment the existing text descriptions with beats and downbeats location (inferred from tempo information in the text prompt), underlying chord progression, key, and tempo as control information. During inference, these additional descriptive texts could steer the music generation towards user-specified music quality. We use state-of-the-art music information retrieval (MIR) methods~\cite{chordino,heydari2021beatnet,bogdanov2013essentia} to extract such control information from our training data. 
Furthermore, to diversify the music samples in the training set, we augment this dataset with variants of the existing music, altered along three aspects---tempo, pitch, and volume---that essentially determine the rhythmic, harmonic, and interpretive aspects of music. The text descriptions are also altered accordingly. The resulting \dataset{} dataset is 11 times the size of the original \texttt{MusicCaps}~\cite{agostinelli2023musiclm} dataset. Our proposed controllable text-to-music model \model{} incorporates a novel MuNet (music-domain-knowledge-informed UNet) that can instill the input chords, beats, key, and tempo, along with the textual description, in the generated music during the reverse-diffusion process. The results in \Cref{sec:experiment} indicate \model{} creates more musically meaningful output and shows improved controllability (e.g., changing chords) over the existing text-to-music models. Our \dataset{} dataset, \modelemoji{} implementation, and comparative music samples are available through \url{https://github.com/AMAAI-Lab/mustango}.



The overall contributions of this paper are:
\begin{enumerate}[label=(\roman*), itemsep=0pt, leftmargin=*, wide, labelwidth=0pt, labelindent=0pt, parsep=0pt, topsep=0.5pt]
  \item We propose \modelemoji{}, a text-to-music diffusion model with our novel MuNet module to explicitly guide the music generation towards input tempo, key, chords, and general textual description. 
  \item We release the \dataset{} dataset with $\sim$53K pairs of music audio and description with information on musical attributes like chords, key, and beats. This is achieved by altering the music samples of \texttt{MusicCaps} along the harmony, tempo, and volume dimensions and enriching the captions with the aforementioned musically-relevant attributes.
  \item We empirically verify that our \model{} model is able to generate high quality music faithful to the input text descriptions, chords, and beats.
\end{enumerate}

\section{Dataset Creation}\label{sec:method}

In this section, the methods of music feature extraction and data augmentation are introduced. Then, the application of these methods and the details of our dataset are discussed.

\subsection{Feature Extraction and Description Enrichment}
\label{sec:method_feature_extract}

We extract four common music features: beats and downbeats, chords, keys, and tempo, and use them to enhance the text prompts and guide music generation. We use BeatNet~\cite{heydari2021beatnet} to extract the beat and downbeat features, $b\in \mathbb{R}^{L_{beats}\times 2}$, where the first dimension represents the type of beat according to the meter (e.g., 1, 2, 3) and the second represents the timing of each corresponding beat in seconds. The second feature \emph{tempo}, measured in beats per minute (BPM), is estimated by averaging the reciprocal of the time interval between beats. Chordino~\cite{chordino} is used to extract the chord features, $c\in \mathbb{R}^{L_{chords}\times 3}$, where the first dimension represents the roots of the chord sequence, the second represents the chord type (e.g., major, minor, maj7, etc.), and the third represents whether the chords are inverted. Finally, Essentia's \cite{bogdanov2013essentia} KeyExtractor algorithm\footnote{\rurl{essentia.upf.edu/reference/std_KeyExtractor.html}} is used to extract the key. The extracted features are used to enrich the textual descriptions and guide the reverse diffusion process. We notice a similar data enrichment approach in concurrent research~\cite{gardner2023llark}.



These features are then expressed in text format following several text templates (e.g., `The song is in the key of A minor. The tempo of this song is Adagio. The beat counts to 4. The chord progression is Am, Cmaj7, G.'). We refer to these as control sentences and they will be appended to the original text prompt to form the enhanced prompts. A full list of the different control sentence templates can be found in \cref{sec:controlcaptions} (\cref{tab:app_attr_to_text}).

\subsection{Augmentation and Music Diversification}\label{sec:method_data_aug}


Our dataset augmentation for both music audio and text prompts increases the total amount of training data 11-fold to improve both audio quality and controllability of our model. Standard text-to-audio augmentations may not suit the nature of music audio. For example, the augmentation method used for \texttt{Tango}~\cite{ghosal2023tango}, whereby two audio samples normalized to similar audio levels are superimposed and their prompts concatenated, would not work for music due to overlapping rhythms, dissonance in harmony, and overall musical concept mismatch.


Therefore, we alter individual music samples along one of the three dimensions---pitch, speed, and volume---which determine the melodic, rhythmic, and dynamic aspects of music. We use \texttt{PyRubberband}\footnote{\rurl{github.com/bmcfee/pyrubberband}} to shift the pitch of the music audio within a range of $\pm$3 semitones following a uniform distribution. We decided to use this range in order to keep the timbre of instruments relatively untouched, as larger pitch shifts could result in unnatural timbre. We change the speed of the music audio by $\pm$(5 to 25)\%, drawn from a uniform distribution as well. Finally, we alter the volume of the audio by introducing a gradual volume change (both crescendo and decrescendo) with the minimum volume drawn from a uniform distribution from 0.1 to 0.5 times the original track's amplitude, while the maximum volume is kept untouched.

The text descriptions are enhanced and modified in tandem with the alterations to the music audio. However, to enhance the robustness of the model, we randomly discard one to four sentences from the prompt that describe the aforementioned music features. More details are illustrated in the Appendix. Finally, we used ChatGPT~\cite{chatgpt} to rephrase the text prompts to add variety to the text prompts.

\subsection{MusicBench}\label{sec:musicBench}
In this study, we make use of the MusicCaps~\cite{agostinelli2023musiclm} dataset, which comprises a collection of 5,521 audio clips featuring music. Each clip is 10 seconds long and is sourced from the train and evaluation splits of the AudioSet~\cite{gemmeke2017audio} dataset. These audio clips are accompanied by on average four-sentence-long English caption that describe the music. However, due to the inaccessibility of some audio files, our dataset comes from 5,479 samples.

\begin{figure}[h]
    \centering
    \includegraphics[width=0.9\linewidth]{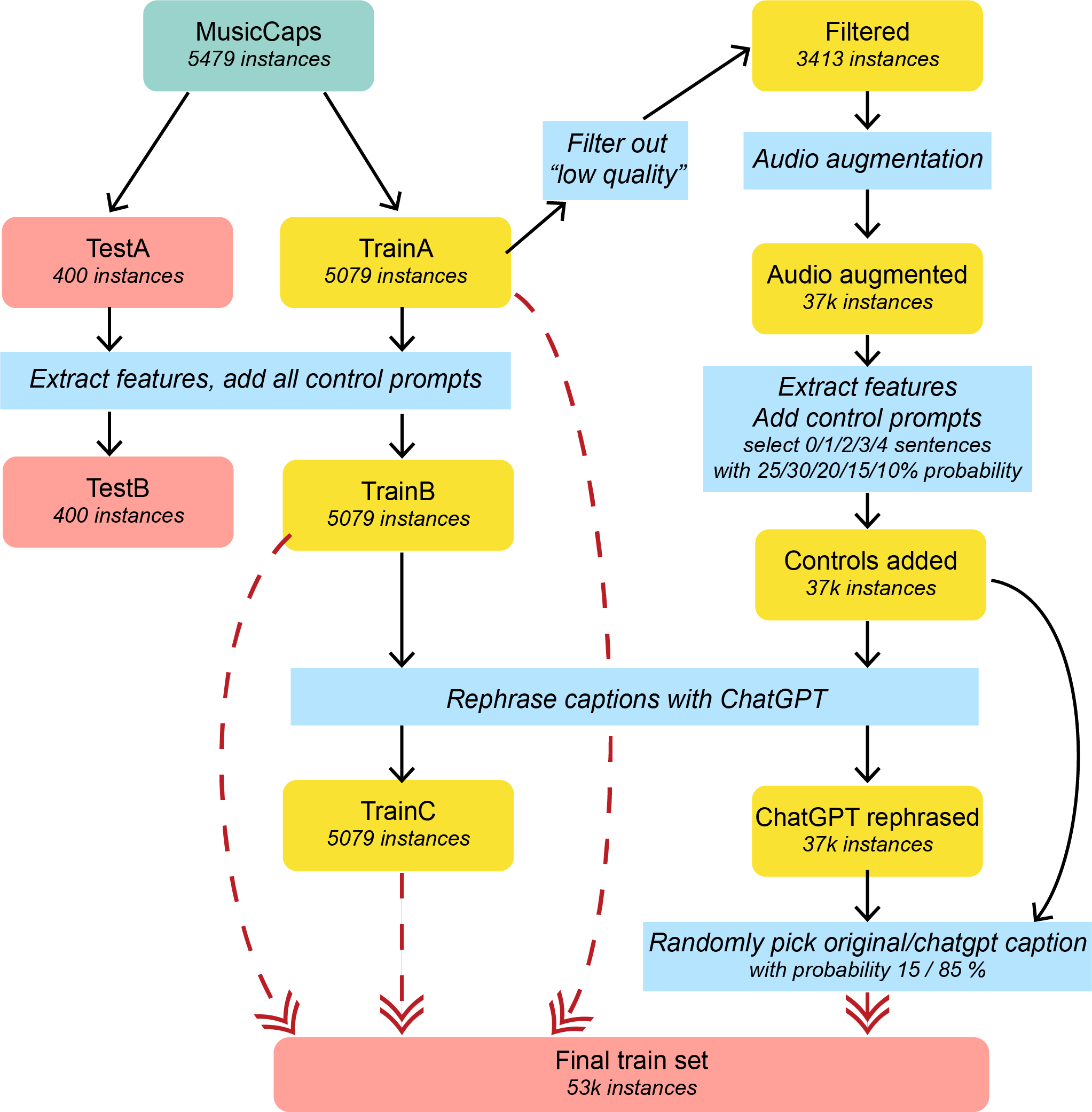}
    \caption{Composition of \dataset{} dataset.}
    \label{fig:data}
\end{figure}

We split our dataset as shown in \cref{fig:data}. First, we split the data into TrainA and TestA sets. 
Subsequently, four control sentences corresponding to the music features are spliced with the original prompts to obtain the TrainB and TestB sets from TrainA and TestA, respectively. Then, by instructing ChatGPT to rephrase the TrainB text prompts, we get the final TrainC set. 

In addition, before performing audio augmentation, we filter out `low quality' samples by removing samples that mention the terms `quality' (as it is typically related to poor quality) or `low fidelity' in the captions of TrainA set, to get 3,413 instances. The higher quality samples are altered (see \cref{sec:method_data_aug}) to form a set of 37k augmentation samples, comprising 6 pitch-shifted, 4 tempo-altered and 1 volume-altered sample per original sample. In the case of pitch-shifted samples, instead of randomly sampling from a uniform distribution, we used all 6 unique semitone shifts (from -3 to +3, excluding 0).
Thereafter we randomly select control prompts to concatenate with the original captions. We pick $0/1/2/3/4$ prompts with a probability of $25/30/20/15/10 \%$, respectively. We do this to increase the robustness of the model, as the model should be able to take inputs both with and without control sentences specifying the four music features. Then, to further increase text input robustness, we rephrase all of the captions using ChatGPT (see \cref{sec:app-chatgpt}). We find this step a necessary addition in our augmentation pipeline as the audio augmentation produces 11 similar samples that share a big portion of their caption with the original MusicCaps caption. By paraphrasing, we create more unique instances. In our final training dataset, we use both of the rephrased and non-rephrased prompts with a probability of $85/15 \%$, respectively. Finally, we take this augmented set and concatenate it with sets TrainA, TrainB, and TrainC to get our final training set consisting of 52,768 samples, hereafter referred to as \dataset{}. We note that TestA and TestB sets consist of 200 `low quality' (as explained above), and 200 `high quality' samples. This means that the test set distribution is slightly different from that of train set. Our intention was to create a difficult evaluation set to test the controllability of \modelemoji{} in tougher conditions.

\section{\modelfig{}}

\begin{figure*}
    \centering
    \includegraphics[width=0.8\textwidth]{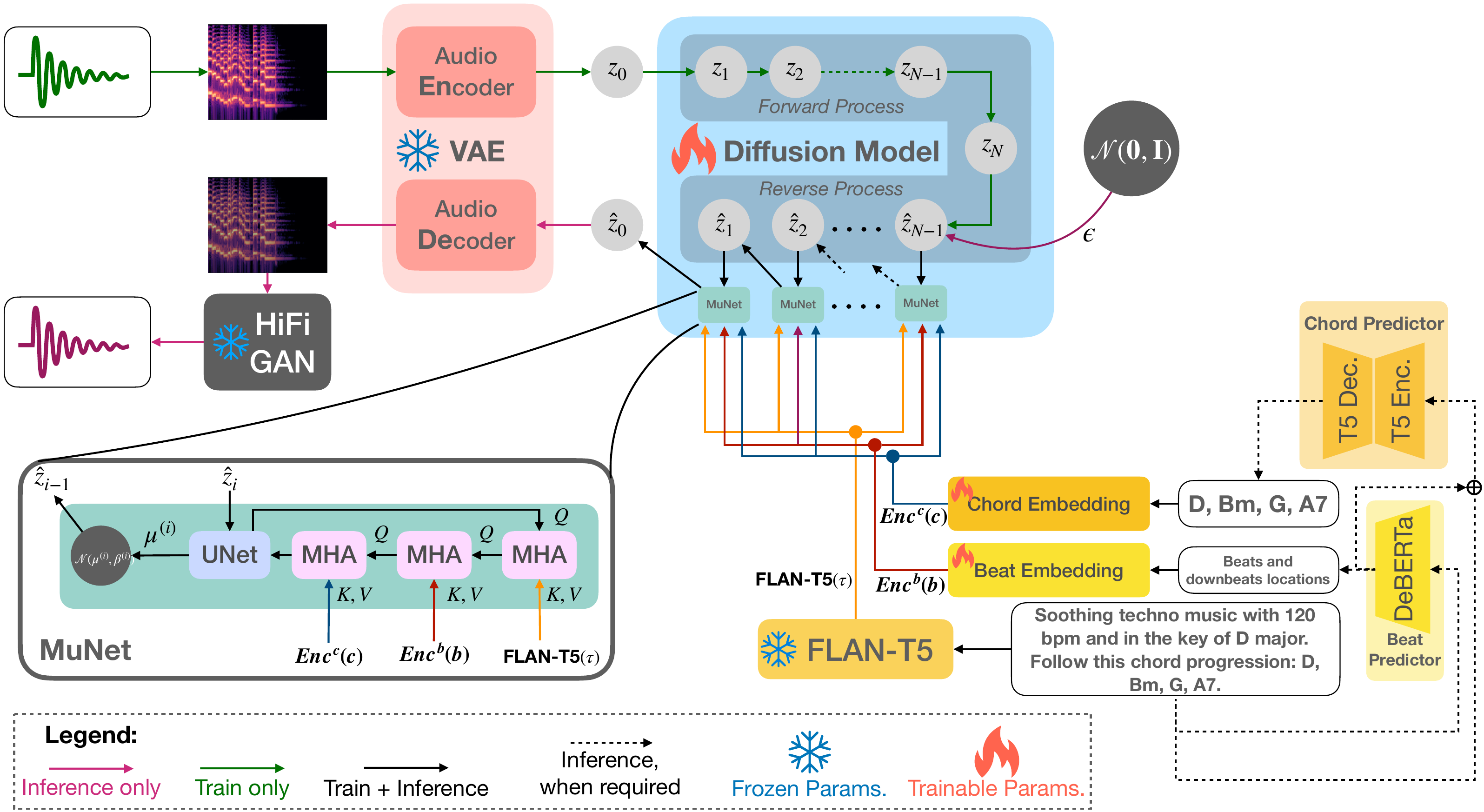}
    \caption{Depiction of our proposed \model{} model. Beats and chords are inferred from the caption when they are not provided as input.}
    \label{fig:model}
\end{figure*}



\model{} consists of two components: 1) \texttt{Latent Diffusion Model}; 2) \texttt{MuNet}.


\subsection{Latent Diffusion Model (LDM)}\label{sec:ldm}

Inspired by \texttt{Tango}~\cite{ghosal2023tango} and \texttt{AudioLDM}~\cite{liu2023audioldm}, we leverage the latent diffusion model (LDM) to reduce computational complexity meanwhile maintaining the expressiveness of the diffusion model. More specifically, we aim to construct the latent audio prior $z_0$ extracted using an extra variational autoencoder (VAE) with condition $\mathcal{C}$ , which in our case refers to a joint music and text condition. Similar to \texttt{Tango}, we leverage the pre-trained VAE from \texttt{AudioLDM} to obtain the latent code of the audio.

Through the forward-diffusion process (Markovian Hierarchical VAE), the latent audio prior $z_0$ turns into a standard gaussian noise $z_N\sim \mathcal{N}(\mathbf{0}, \mathbf{I})$, as shown in \cref{eq:forward_diff} where a pre-scheduled gaussian noise ($0<\beta_1<\beta_2<\dots<\beta_N<1$) is gradually added at each forward step:
\begin{flalign}
q(z_n|z_{n-1}) &= \mathcal{N}(\sqrt{1-\beta_n} z_{n-1}, \beta_n \mathbf{I}).  \label{eq:forward_diff}
\end{flalign}

For the reverse process, which reconstructs $z_0$ from Gaussian noise $z_N\sim \mathcal{N}(\mathbf{0}, \mathbf{I})$, we propose \texttt{MuNet} (see \cref{sec:munet}), which is able to steer the generated music towards the given condition $\mathcal{C}$. Intuitively, backward diffusion aims to iteratively reconstruct the latent audio prior $z_{n-1}$ from the previous step $z_n$ until $z_0$, using a denoiser $\hat\epsilon_\theta^{(n)}(z_n, \mathcal{C} )$. This denoiser is driven by classifier-free guidance, similar to \texttt{Tango}. The reverse diffusion process is outlined in \cref{sec:rev_diff}.


This reconstruction is trained using a noise-estimation loss where $\hat\epsilon_\theta^{(n)}$ is the estimated noise and $\gamma_n$ is the weight of reverse step $n$:
\setlength{\abovedisplayskip}{3pt}
\setlength{\belowdisplayskip}{4pt}
\begin{flalign*}
    \mathcal{L}_{DM} = \sum_{n=1}^N\gamma_n \mathbb{E}_{\epsilon_n\sim \mathcal{N}(\mathbf{0}, \mathbf{I}), z_0} || \epsilon_n -  \hat\epsilon_\theta^{(n)}(z_n, \mathcal{C} ) ||_2^2.
\end{flalign*}

\subsection{\texttt{MuNet}}
\label{sec:munet}

The reverse-diffusion process, briefly described in \cref{sec:ldm}, is conditioned on both musical attributes (beat $b$ and chord $c$) and text $\tau$ ($\mathcal{C}:=\{\tau, b, c\}$). This is realized through the Music-Domain-Knowledge-Informed UNet (MuNet) denoiser as follows:


\setlength{\abovedisplayskip}{2pt}
\setlength{\belowdisplayskip}{9pt}

\begin{flalign}
    & U^{(1)} = z_n \nonumber \\
    & A_\tau =\text{MHA}_{\theta_\tau}(Q=U^{(l)}, K/V=\text{FLAN-T5}(\tau)) \nonumber \\ 
    & A_b =\text{MHA}_{\theta_b}(Q=A_\tau, K/V=\boldsymbol{Enc^b(b)}) \nonumber \\
    & A_c =\text{MHA}_{\theta_c}(Q=A_b, K/V=\boldsymbol{Enc^c(c)}) \nonumber \\
    & U^{(l+1)} = \text{UNet}^{(l)}_\theta(A_c) \nonumber \\
    & \epsilon_\theta^{(n)}(z_n, \mathcal{C}) := U^{(L+1)}
    \label{eqn:noise-estimator}
\end{flalign}
where, MHA is the multi-headed attention block~\cite{vaswani2017attention} for the cross attentions, where $Q, K,$ and $V$ are query, key, and value, respectively, and FLAN-T5 is the text encoder model~\cite{chung2022scaling}, adopted from \texttt{Tango}. 
We prioritize applying cross-attention to the beat first, as we consider a consistent rhythm to be the fundamental basis for the generated music. Subsequently, we can focus on conditioning based on chords.

MuNet consists of a UNet~\cite{UNET}---consisting of in total $L$ downsampling, middle, and upsampling blocks---and multiple conditioning cross-attention blocks. We use two encoders, $\boldsymbol{Enc^b}$ and $\boldsymbol{Enc^c}$, to encode the beat and chord features which leverage both the state-of-the-art Fundamental Music Embedding (FME) as well as an onset-and-beat-based positional encoding~\cite{FME} which we name Music Positional Encoding (MPE). These ensure the musical features are properly captured and several fundamental music properties (e.g., intervals between pitches are translational invariant) are preserved. 

We introduce the two encoders $\boldsymbol{Enc^b}$ and $\boldsymbol{Enc^c}$ that extract the beat and chord embeddings from the raw input. The beat encoder $\boldsymbol{Enc^b}$, defined in \cref{eq:beat_enc}, encodes the beat types $b[:, 0]$ (\cref{sec:method_feature_extract}) using One-Hot Encoding ($\boldsymbol{OH_b}$) and the beat timings $b[:, 1]$ with Music Positional Embedding. By concatenating these beat types and timing encodings and passing them through a trainable linear layer ($\boldsymbol{W_b}$), we obtain the final beat features:

\begin{flalign}
    \boldsymbol{Enc^b}(b) &:= \boldsymbol{W_b} (OH_b(b[:, 0])\oplus MPE(b[:, 1]))\label{eq:beat_enc} \\
\boldsymbol{Enc^c}(c) &:= \boldsymbol{W_c} (\text{FME}(c[:, 0]) \oplus OH_t(c[:, 1]) \oplus \nonumber \\
&OH_i(c[:, 2]) \oplus \text{MPE}(c[:, 3]))\label{eq:chord_enc}
\end{flalign}

In the chord encoder in \cref{eq:chord_enc}, we obtain the chord embeddings by first concatenating i) FME-embedded~\cite{FME} chord roots $c[:, 0]$ (see \cref{sec:method_feature_extract}); ii) One-Hot encoded chord type ($c[:, 1]$); iii) One-Hot encoded chord inversions ($c[:, 1]$); and iv) MPE-embedded~\cite{FME} timing of the chords ($c[:, 3]$). Subsequently, this concatenated representation is passed through a trainable linear layer ($\boldsymbol{W_c}$). Notably, we incorporate a music-domain-knowledge informed music embedding through the use of the Fundamental Music Embedding from ~\citet{FME}, which effectively captures the translational invariant property of pitches and intervals, resulting in a more musically meaningful representation of the chord. 

After obtaining the encoded beat and chord embeddings, we use two additional cross-attention layers to integrate these music conditions during the denoising process, whereas \texttt{Tango} used one cross-attention layer to incorporate only text conditions. This enables MuNet to leverage both music and text features during the denoising process, resulting in more controllable and meaningful music generation.

\subsection{Inference}
During the training phase, we use teacher forcing and hence utilize the ground truth beats and chord features to condition the music generation process. However, during inference, we employ two transformer-based text-to-music-feature generators that have been trained independently to predict the beat and chord features as follows: 

\textbf{Beats}: We use the DeBERTa Large model ~\cite{he2022debertav3} as the beats predictor. The model takes the text caption as input and predicts: i) the beat count (meter) of corresponding music, and ii) the sequence of interval duration between the beats. We predict them from the token representations of the final layer of the model. The beat count takes an integer value between 1 and 4 for the instances in our training data. Hence, we predict the beat using a four-class classification setup from the first token of the output layer. The interval durations are predicted as a float value from the second token onwards. As an example, if the beat count is predicted as $2$ and the interval durations are predicted as $t_1, t_2, t_3, \dots$, then the predicted beats are as follows: $1$ at $t_1$, $2$ at $t_1 + t_2$, $1$ at $t_1 + t_2 + t_3$, etc. We keep the predicted beats time up to 10 seconds and ignore predicted timestamps beyond that. 

\textbf{Chords}: We use the sequence to sequence FLAN-T5 Large model ~\cite{chung2022scaling} as the chords predictor. The model takes the concatenation of the text caption and the verbalized beats as input. The verbalized beats are prepared for the example we illustrated earlier as follows: \textit{Timestamps: $t_1$, $t_1+t_2$, $t_1+t_2+t_3 \dots$, Max Beat: $2$}. The model is trained to generate the verbalized chords sequence with timestamps, which would look like something as follows: \textit{Am at 1.11; E at 4.14; C\#maj7 at 7.18}. We again keep the predicted chord time up to 10 seconds and ignore timestamps predicted beyond that. 



\section{Experiments}
\label{sec:experiment}
We conduct extensive objective and subjective evaluations to answer these research questions: 
i) How is the audio quality of the music generated by \modelemoji{}? ii) Does \modelemoji{} generate music with better music quality compared to other baselines? iii) Is \modelemoji{} more controllable in terms of music-specific instructions? iv) Is our data augmentation approach effective -- can models trained on only this dataset compete with large-scale pre-trained models?

\subsection{Baselines and \modelemoji{} Variants}\label{sec:baselines_models}


We first compare \modelemoji{} with \texttt{Tango} since it shares a similar architecture with \modelemoji{}, except for the extra conditioning module: MuNet. 
To judge the efficacy of \modelemoji{}, we train the following three models from scratch:
i) \texttt{Tango} trained on MusicCaps TrainA, ii) \texttt{Tango} trained on \dataset{}, iii) \modelemoji{} trained on \dataset{}. Additionally, we finetune \texttt{Tango} and \modelemoji{} from pre-trained \texttt{Tango} checkpoints: iv) pre-trained \texttt{Tango} 
fine-tuned on AudioCaps and MusicCaps\footnote{\rurl{hf.co/declare-lab/tango-full-ft-audio-music-caps}}, v) pre-trained \texttt{Tango} fine-tuned on AudioCaps\footnote{\rurl{hf.co/declare-lab/tango-full-ft-audiocaps}}, then finetuned on \dataset{}, vi) \modelemoji{} initialized from pre-trained \texttt{Tango} and finetuned on \dataset{}. 
Furthermore, we compare \modelemoji{} with state-of-the-art Text-to-Music model of \texttt{MusicGen}~\cite{copet2023simple} and a Text-to-Audio model of \texttt{AudioLDM2}~\cite{liu2023audioldm2}. For \texttt{MusicGen} baselines, we use the small and medium checkpoints. For \texttt{AudioLDM2}, we compare with their music-specific checkpoint.

\begin{table*}[t!]
\centering
\small
\resizebox{\textwidth}{!}{
\begin{tabular}{lccc|rrr|rrr|rrr}
\toprule
\multirow{2}{*}{\textbf{Model}} & \multirow{2}{*}{\textbf{Datasets}} & \multirow{2}{*}{\textbf{Pre-trained}} & \multirow{2}{*}{\textbf{\#Params}} & \multicolumn{3}{c}{\textbf{TestA}} & \multicolumn{3}{|c}{\textbf{TestB}} & \multicolumn{3}{|c}{\textbf{\datasetFMA{}}} \\ 
& & & & FD~$\downarrow$ & FAD~$\downarrow$ & KL~$\downarrow$ & FD~$\downarrow$ & FAD~$\downarrow$ & KL~$\downarrow$ & FD~$\downarrow$ & FAD~$\downarrow$ & KL~$\downarrow$ \\
\midrule
\texttt{MusicGen-S} & -- & \redcross & 300M   & 35.40 & 6.82  & 1.81 & 36.40 & 7.54 & 1.75 & 23.21 & 5.13 & 1.31 \\
\texttt{MusicGen-M} & -- & \redcross & 1.5B   & 36.49 & 6.98  & 1.71 & 35.54 & 6.99 & 1.71 & 22.61 & 5.01 & 1.33 \\
\texttt{AudioLDM2} & -- & \redcross & 346M   & 32.76 & 5.29  & 1.68 & 33.66 & 5.42 & 1.75 & \textbf{19.99} & 3.01 & 1.33 \\
\midrule
\texttt{Tango} & \texttt{MusicCaps} & 
\redcross & $866$M & 30.80 & 2.84 & 1.34 & 30.39 & 2.92 & 1.33 & 28.32 & 3.75 & 1.22 \\
\texttt{Tango} & \texttt{MusicCaps} & 
\greencheck & $866$M & 34.87 & 4.05 & 1.25 & 37.85 & 4.52 & 1.32 & 28.81 & 2.92 & 1.21 \\
\texttt{Tango} & \dataset{} & 
\redcross & $866$M & 28.50 & 2.29 & 1.33 & 28.27 & 2.17 & 1.32 & 26.31 & \bf 2.31 & 1.16 \\
\texttt{Tango} & \dataset{} & 
\greencheck & $866$M & \textbf{25.38} & 1.91 & \bf 1.19 & \textbf{24.60} & 1.77 & 1.13 & 24.48 & 2.96 & \bf 1.15 \\
\modelemoji{} & \dataset{} & 
\redcross & $1.4$B & 26.58 & 2.09 & 1.21 & 25.24 & \textbf{1.57} & 1.18 & 24.24 & 2.94 & 1.16 \\
\modelemoji{} & \dataset{} & 
\greencheck & $1.4$B & 26.35 & \textbf{1.46} & 1.21 & 25.97 & 1.67 & \textbf{1.12} & 25.18 & 2.34 & 1.16 \\
\bottomrule
\end{tabular}
}
\caption{Objective evaluation results of the models on TestA, TestB, and \datasetFMA{} datasets. 
}
\label{tab:obj_eval}
\end{table*}


\subsection{Training and Additional Evaluation Set}
\label{sec:trainsetup}
All the models were trained at a learning rate of $4.5e-5$ using the AdamW~\cite{loshchilov2017decoupled} optimizer until convergence.
Our Beat and Chord predictors are also trained on \dataset{}. More details on training the classifier-free guidance, and parameters are reported in \cref{sec:app-training}.

Given that some of the fine-tuned models used in our experiments were exposed to the entire MusicCaps dataset in the initial \texttt{Tango} pre-trained checkpoint, we can only fairly evaluate those models on a different and independently created evaluation set. We thus curated 1,000 pseudo-captioned evaluation samples from the music files of Free Music Archive (FMA) \cite{defferrard2016fma}, which we refer to as \datasetFMA{}. The details of creating \datasetFMA{} are reported in \cref{sec:fmacaps}.

\subsection{Inference Settings and Time}
\label{sec:inf_time}
In all our experiments, we use 200 diffusion steps with a classifier-free guidance scale of three for all variants of \modelemoji{}, \texttt{Tango}, and \texttt{AudioLDM2}. In \texttt{MusicGen}, we generate audio sequences of 10 seconds to match the outputs of the other models.
We further performed a simple time measurement experiment to assess computing time of presented models. With batch size of 1, we inferred 20 samples and took the average inference time on a single Tesla V100 GPU. The obtained results are: \texttt{Tango} 34 sec, \texttt{MusicGen-M} 51 sec, \modelemoji{} 76 sec.

\subsection{Objective Evaluation Methodology}
\label{sec:objeval}

\paragraph{Audio Quality Estimation}
The quality of the generated audio samples is evaluated using three objective metrics: Fréchet Distance (FD), Fréchet Audio Distance (FAD) \cite{kilgour2019frechet}, and Kullback-Leibler divergence (KL) as used earlier in \texttt{AudioLDM} and \texttt{Tango}. 

\paragraph{Controllability Evaluation}

We evaluate each model's controllability using TestB (see \cref{sec:musicBench}) and a version of \datasetFMA{} that has all the control sentences for each sample in the prompt. We first generate music based on the text prompts and then extract the musical features mentioned in \cref{sec:method_feature_extract}. Subsequently, we define nine metrics (all represented in percentage; in the case of binary values, 100 stands for true and 0 stands for false) to evaluate whether the music properties in the generated music match the text prompts. The metrics are:

\begin{itemize}[itemsep=0pt, leftmargin=*, wide, labelwidth=0pt, labelindent=0pt, parsep=0pt]
    \item \textbf{Tempo Bin (TB)}: The predicted beats per minute (bpm) fall into the ground truth tempo bin.
    \item \textbf{Tempo Bin with Tolerance (TBT)}: The predicted bpm falls into the ground truth tempo bin or a neighboring one.
    \item \textbf{Correct Key (CK)}: The predicted key matches the ground truth key. 
    \item \textbf{Correct Key with Duplicates (CKD)}: The predicted key matches the ground truth key or an equivalent key (i.e., a major key and its relative minor). 
    \item \textbf{Perfect Chord Match (PCM)}: The predicted chord sequence perfectly matches ground truth in terms of length, order, chord root, and chord type. 
    \item \textbf{Exact Chord Match (ECM)}: The predicted chord sequence matches the ground truth exactly in terms of order, chord root, and chord type, with tolerance for missing and excess chord instances. 
    \item \textbf{Chord Match in any Order (CMO)}: The portion of predicted chord sequence matching the ground truth chord root and type, in any order. 
    \item \textbf{Chord Match in any Order major/minor Type (CMOT)}: The portion of predicted chord sequence matching the ground truth in terms of chord root and binary major/minor chord type, in any order (e.g., D, D6, D7, Dmaj7 are all considered major). 
    \item \textbf{Beat Match (BM)}: The percentage of predicted beat counts that match the ground truth. 
\end{itemize}

\begin{table*}[t!]
\centering
\resizebox{\textwidth}{!}{
\begin{tabular}{lcc|rr|rr|rrrr|r|rr|rr|rrrr|r}
\toprule
\multirow{3}{*}{\textbf{Model}} & \multirow{3}{*}{\textbf{Datasets}} & \multirow{3}{*}{\begin{tabular}{c}\textbf{Pre-}\\\textbf{trained}\end{tabular}} & \multicolumn{9}{c}{\textbf{TestB}} & \multicolumn{9}{|c}{\textbf{\datasetFMA{}}} \\ 
\cmidrule{4-21}
& & & \multicolumn{2}{c}{\textbf{Tempo}} & \multicolumn{2}{|c}{\textbf{Key}} & \multicolumn{4}{|c}{\textbf{Chord}} & \multicolumn{1}{|c}{\textbf{Beat}} & \multicolumn{2}{|c}{\textbf{Tempo}} & \multicolumn{2}{|c}{\textbf{Key}} & \multicolumn{4}{|c}{\textbf{Chord}} & \multicolumn{1}{|c}{\textbf{Beat}} \\
& & & TB & TBT & CK & CKD & PCM & ECM & CMO & CMOT & BM & TB & TBT & CK & CKD & PCM & ECM & CMO & CMOT & BM \\
\midrule
\texttt{MusicGen-S} & -- &
\redcross & 39.50 & 56.00  & 17.5 & 19.00  & 3.17  & 6.03  & 12.56 & 21.74  & 36.75 & \textbf{45.0} & 61.9 & 19.9 & 21.1 & 3.62  & 6.49  & 10.99 & 22.30  & 42.4 \\
\texttt{MusicGen-M} & -- &
\redcross & \textbf{41.00} & \textbf{60.25} & 25.5 & 26.25 & 3.97  & 8.21  & 14.42 & 26.76  & \textbf{45.00}  & 42.7 & \textbf{63.5} & 23.5 & 24.3 & 6.38  & 10.60 & 16.24 & 31.51  & \bf 42.9 \\
\texttt{AudioLDM2} & -- &
\redcross & 21.25 & 47.75 & 6.50 & 10.25 &  0.79 & 2.67  & 4.87  & 10.55 & 39.75 & 24.2 & 48.7 & 5.9  & 9.9 & 1.06 & 1.96  & 3.27  & 8.84  & 38.1 \\
\midrule
\texttt{Tango} & \texttt{MusicCaps} & 
\redcross & 26.00 & 55.25 & 4.00 & 7.00 & 0.53 & 2.09 & 4.30 & 11.13 & 41.00 & 22.5 & 49.6 & 3.6 & 8.6 & 0.64 & 1.43 & 4.03 & 10.82 & 41.1 \\
\texttt{Tango} & \texttt{MusicCaps} & 
\greencheck & 27.50 & 52.00 & 7.75 & 11.25 & 1.06 & 3.07 & 6.72 & 13.99 & 36.75 & 24.2 & 48.6 & 5.9 & 8.6 & 1.17 & 2.74 & 5.17 & 12.69 & 35.4  \\
\texttt{Tango} & \dataset{} & 
\redcross & 24.75 & 50.75 & 34.25 & 34.50 & 5.56 & 12.03 & 21.54& 32.21& 34.25 & 25.5 & 51.0 & \bf 38.1 & \bf 38.4 & 6.60 & 13.45 & 21.18 & 41.49 & 36.4 \\
\texttt{Tango} & \dataset{} & 
\greencheck & 26.00 & 48.75 & 30.25 & 31.00 & 6.61 & 13.33 & 22.53 & 39.31 & 38.50 & 22.8 & 45.6 & 30.6 & 31.7 & 7.55 & 14.72 & 22.35 & 44.46 & 36.0 \\
\modelemoji{} & \dataset{} & 
\redcross & 25.50 & 52.00 & \textbf{41.75} & \textbf{42.50} & \textbf{17.99} & \textbf{32.61} & \textbf{48.74} & \textbf{68.46} & 42.00 & 24.1 & 50.9 & 36.8 & 37.3 & \textbf{23.94} & \textbf{35.43} & \textbf{49.59} & \textbf{75.83} & 42.6 \\
\modelemoji{} & \dataset{} & 
\greencheck & 21.25 & 48.25 & 34.50 & 35.50 & 11.64 & 20.82 & 32.93 & 50.56 & 34.75 & 26.2 & 52.2 & 33.9 & 34.7 & 15.21 & 25.48 & 37.50 & 61.55 & 39.1 \\
\bottomrule
\end{tabular}
}
\caption{Controllability evaluation results 
of the models on TestB and full-control variant of \datasetFMA{}. 
Higher numbers indicate better controllability.  
}
\label{tab:control_eval}
\end{table*}

\subsection{Objective Evaluation Results}\label{sec:obj_eval_results}

\paragraph{Audio Quality:} The results for TestA, TestB and \datasetFMA{} are presented in \Cref{tab:obj_eval}. Both \texttt{Tango} variants trained on MusicCaps are inferior to the other four models, which depicts the efficacy of our augmentation strategy. Pre-trained \texttt{Tango} fine-tuned on \dataset{} and \modelemoji{} pre-trained seem to perform very similarly in FD and KL, but \modelemoji{} pre-trained shows a big improvement in FAD, which suggests better-perceived quality and musicality as FAD is a human perception-inspired metric. Lastly, the performance of \modelemoji{} trained from scratch is comparable in FD and KL to both pre-trained versions of \modelemoji{} and \texttt{Tango} trained on \dataset{}, which shows that training with our augmented dataset can be an alternative to large-scale audio pre-training for music generation. \modelemoji{} also outperforms \texttt{MusicGen} and \texttt{AudioLDM2} in FAD and KL across all three sets.

We note that the results for \texttt{MusicGen} and \texttt{AudioLDM2} differ from what was reported in their original papers in evaluation on MusicCaps. This is due to \dataset{} representing a different, more challenging, split of data than the MusicCaps evaluation set, as described in \cref{sec:musicBench}. Additionally, we note that the results of both \texttt{MusicGen} and \texttt{AudioLDM2} show more improvement than \modelemoji{} when evaluated on \datasetFMA{} as compared to TestA and TestB. This is due to the fact that \modelemoji{} was trained on \dataset{}, thus TestA and TestB represent similar distributions to the training set, while \datasetFMA{} is of a slightly different distribution. The \texttt{MusicGen} and \texttt{AudioLDM2} models on the other hand, were trained with various large-scale data, hence they perform well on an unseen set. 

\paragraph{Controllability:} The evaluation results on controllability are shown in \Cref{tab:control_eval}. On TestB, in terms of Tempo metrics, all the models perform comparably, except for \texttt{MusicGen}, which performs better. In Beat metrics, the models perform similarly to each other. \modelemoji{} placed second closely behind MusicGen.
The similarity in performance among the models could be caused by the MusicCaps dataset already containing enough information about tempo, with words such as "slow", "fast", "moderate", etc. This information being passed through the text encoding might be a sufficient control command.
Furthermore, the inaccuracy of the beat extractor combined with the fact that not all music pieces in MusicCaps have clearly audible beats might further contribute to the Beat and Tempo metrics results. Thus, having more open-sourced high-quality music data would greatly benefit development of even more controllable systems.

In Key metrics, we can observe that models trained on \dataset{} perform significantly better than the ones trained on MusicCaps. Additionally, \modelemoji{} outperforms all the other models on TestB and placed second on \datasetFMA{}. Finally, in Chord controllability, \modelemoji{} outperforms all the other models by a big margin. On \datasetFMA{}, we further see that the Chord metrics are even better for \modelemoji{} with CMOT reaching 75.83. Overall, the results gathered from both TestB and modified \datasetFMA{} correlate in most aspects. Overall, \modelemoji{} performs fine in Beat and Tempo metrics, and it excels in Key and Chord controllability.

\subsection{Subjective Evaluation Methodology}
\label{sec:subjeval}

We conducted two rounds of subjective evaluation, each consisting of a general and an expert listening test that focuses on controllability. The first round is aimed at comparing \modelemoji{} variants with \texttt{Tango} and in the second run we compare \modelemoji{} with the state-of-the-art models: \texttt{MusicGen} and \texttt{AudioLDM2}.

In the first round of the general listening test, subjects listened to ten generated music samples for each of the four models (pre-trained \modelemoji{}, \modelemoji{}, \texttt{Tango} trained with MusicCaps and MusicBench) and were provided with the input text caption. The ten text prompts were custom-made by music experts in the style of MusicCaps, and are shown in \Cref{tab:controlprompts} in the Appendix. The participants were asked to rate the: i) audio rendering quality (AQ), ii) relevance of the audio with the input text prompt (REL), iii) overall musical quality (OMQ), iv) rhythm consistency (RC), and v) harmony and consonance of music (HC).
For the expert listening test, we added two additional music control-specific aspects to rate the degree to which the chords and tempo from the generated music match the text prompt. We denote them as MCM and MTM (musical chord/tempo match).
All the aspects were rated on a 7-point Likert scale using the PsyToolkit interface \cite{stoet2010psytoolkit}. The full questions and interface used are shown in \cref{sec:eval-interface}.  

For the expert listening test in the first round, we found experts with at least five years of formal musical training who can identify music attributes from music audio. They were presented with 80 samples generated using 20 custom text prompts for each of the four models as shown in \Cref{sec:captions}. Samples consisted of ten \emph{contrasting} pairs (e.g., same prompts with different chord changes) that aimed to target musical controllability.


In the second round of the general listening test, we used the same 10 captions as in the first run with five additional captions taken from FMACaps. In the expert test, we kept the same 10 contrasting pairs as in the first round. For both of these tests, we downsampled the \texttt{MusicGen} samples to 16 kHz to eliminate audio quality bias in listeners' responses, and we excluded the AQ metric from these tests.

\subsection{Subjective Evaluation Results}\label{sec:sub_eval_results}

\begin{table*}[t!]
    \centering
    \begin{adjustbox}{width=1\linewidth,center}
    \begin{tabular}{lcc|ccccc|ccccccc}
    \toprule
    \multirow{2}{*}{\textbf{Model}} & \multirow{2}{*}{\textbf{Datasets}} & \multirow{2}{*}{\textbf{Pre-trained}} & \multicolumn{5}{c}{\textbf{General audience}} & \multicolumn{7}{|c}{\textbf{Music experts}} \\
    & & & REL & AQ & OMQ & RC & HC & REL & MCM & MTM & AQ & OMQ & RC & HC \\
    \midrule
     \texttt{Tango} & \texttt{MusicCaps} & \greencheck & 4.09 & 3.68 & 3.55 & 3.91 & 3.80 &  4.35 & 2.75 & 3.88 & 3.35 & 2.83 & 3.95 & 3.84 \\
     \texttt{Tango} & \dataset{} & \greencheck & \textbf{4.96} & \textbf{4.26} & \textbf{4.40} & 4.49 & 4.61  & 4.91 & 3.61 & 3.86 & 3.88 & 3.54 & 4.01 & 4.34 \\
     \modelemoji{} & \dataset{} & \greencheck & 4.85 & 4.10 & 4.02 & 4.24 & 4.43 & 5.49 & 5.76 & 4.98 & 4.30 & 4.28 & 4.65 & 5.18 \\
     \modelemoji{} & \dataset{} & \redcross & 4.79 & 4.20 & 4.23 & \textbf{4.51} & \textbf{4.63} & \textbf{5.75} & \textbf{6.06} & \textbf{5.11} & \textbf{4.80} & \textbf{4.80} & \textbf{4.75} & \textbf{5.59} \\
     \midrule
      \texttt{MusicGen-M}  & - & - & 4.55 & - & \bf 4.40 & \bf 5.11 & \bf 4.63  &  4.41 & 2.99 & 4.83 & - & \bf 5.01 & \bf 5.61 & \bf 5.31  \\
     \texttt{AudioLDM2} & - & - & 3.99 & - & 3.89 & 4.38 & 4.11  & 3.71 & 2.48 & 3.53 & - & 3.29 & 3.84 & 3.40 \\
     \modelemoji{}  & \dataset{} & \redcross & \bf 5.18 & - & 4.15 & 4.31 & 4.47  & \bf 5.79 & \bf 6.10 & \bf 4.84 & - & 4.53 & 4.14 & 5.11 \\
    \bottomrule
    \end{tabular}
        \end{adjustbox}
    \caption{Average ratings for each metric in the general and expert listening study. Top part of the table shows the first run of listening tests, the bottom part represents the second round of comparison with \texttt{MusicGen} and \texttt{AudioLDM2}.}
    
    \label{tab:subj_eval}
\end{table*}

A total of 48 participants participated in the first round of the general listening test, of which 26 had more than five years of formal musical training. The results in \Cref{tab:subj_eval} show the average ratings for each of the metrics defined above. We can clearly see that the \texttt{Tango} baseline model is outperformed in all metrics by the models trained on \dataset{}. Interestingly, \modelemoji{} trained from scratch performs the best in terms of audio quality, rhythm presence, and harmony. The differences in ratings are minimal between the three top models, clearly confirming that our augmentation method is effective in furthering the output quality and that \modelemoji{} is able to reach state-of-the-art quality. 

A total of four experts participated in the controllability listening study. The results of the expert listening study in \Cref{tab:subj_eval} further confirm that both \modelemoji{} models outperform the \texttt{Tango} baselines in all metrics, especially in terms of the chords of the generated music matching with the input text caption (Chord Match or MCM). This further supports the controllability results presented in \Cref{tab:control_eval} and shows that our proposed \modelemoji{} model can indeed understand music-specific text prompts. 


In the second run, a total of 17 general audience listeners and 4 experts participated. The results are depicted in the lower part of \cref{tab:subj_eval}. We performed a series of paired t-tests on the obtained results and conclude that \modelemoji{} outperforms \texttt{MusicGen} and \texttt{AudioLDM2} in terms of REL, with a statistically significant difference; and performs similarly in OMQ, HC, and MTM to \texttt{MusicGen}, where the t-tests showed no stastically significant differences (both in general audience and expert test). Moreover, \modelemoji{} dominates in MCM. In RC, \texttt{MusicGen} outperformed both \texttt{AudioLDM2} and \modelemoji{}.

\subsection{Ablation Study}

Although without explicitly ablating one module at a time due to resource constraints, we are able to answer the following research questions:

\paragraph{Is Pre-training \modelemoji{} Necessary?}
 In one of the experiment settings in \cref{sec:baselines_models}, we initialized \modelemoji{} with a pre-trained \texttt{Tango} checkpoint and subsequently fine-tune it using the \texttt{AudioCaps} dataset. This \texttt{Tango} model was pre-trained using 1.2 million text-audio paired samples and it encapsulates a broad understanding of general audio and text. However, we observed this did not prove beneficial for music generation (see \crefrange{tab:obj_eval}{tab:subj_eval}). Nevertheless, these checkpoints may find utility in composing music with soundscapes, such as ``Hip-hop music with a lion's roar in the background.''

\paragraph{Is \texttt{MuNet} Helpful?}
We prove the effectiveness of \texttt{MuNet} in \Cref{sec:obj_eval_results} and \Cref{sec:sub_eval_results}, we show that the use of \texttt{MuNet} significantly enhances the performance of \modelemoji{} in terms of controllability under both objective and subjective evaluations. Moreover, several objective metrics which are not explicitly targeted at controllability (i.e., FD, FAD, and KL-divergence), consistently show superior performance when \texttt{MuNet} is incorporated. With classier-free guidance, \texttt{MuNet} does not compromise the overall quality of the generated music when the control sentences in the prompts are absent. 



\subsection{Discussions}

As both objective and subjective evaluation results show, \modelemoji{} gives state-of-the-art performance in music quality and drastic improvement in music controllability, despite being trained on a publicly available dataset of relatively small size as compared to other available text-to-music systems such as \texttt{MusicGen}, which are usually trained on private large-scale licensed dataset. Although these text-to-music systems usually generate music with better audio quality or longer-term structure, which sheds light on further improvement direction of \modelemoji{}. 

\section{Conclusion}
\label{sec:conclusion}


In conclusion, \modelemoji{} presents a significant advancement in the field of controllable text-to-music generation.  \modelemoji{} is a controllable diffusion-based text-to-music system inspired by music-domain knowledge which is able to generate music that follows certain music properties embedded within user-specified text prompts.  The integration of the MuNet module within \modelemoji{}  enables greater music controllability over state-of-the-art text-to-music systems such as \texttt{Tango}, \texttt{AudioLDM2} and \texttt{MusicGen}. We also made our dataset \dataset{} and model publicly available. \dataset{} contains 11 times more data than the original \texttt{MusicCaps} dataset and includes text prompts that contain music-theory-based description and augmented music audio.




\section{Limitations}


Our music generation method is limited to Western music in terms of controllability since the control information mentioned in the paper (e.g., chord, key) might be missing or appear in a different form in other non-Western music (e.g., Indian or Chinese classical music). We also assumed the availability of paired text captions of music, which was used to train our model. \modelemoji{} is also currently limited to generating music of up to ten seconds due to computational constraints. Adapting \modelemoji{} for generating long-form music is left for future work.

\section{Ethical Considerations}
Our training data is based on the MusicCaps dataset~\cite{agostinelli2023musiclm}. The 5.5k music samples in Music-Caps are sourced from Youtube under Creative Commons license. We perform our custom data augmentation strategies solely on this dataset. We did not use any other privately-licensed dataset. 

Our listening tests involved human annotators for which the data collection protocol was approved by an independent ethics review board. More details can be found in the \Cref{sec:eval-interface}.

\section*{Acknowledgements}
This project has received funding from SUTD MOE grant SKI 2021\_04\_06 and AcRF MoE Tier-2 grant (Project no. T2MOE2008, and Grantor reference no. MOE-T2EP20220-0017) titled: “CSK NLP: Leveraging Commonsense Knowledge for NLP”, for the support. This work is also supported by the Microsoft Research Accelerate Foundation Models Academic Research program. We also thank Huggingface for sponsoring the live demo on Huggingface Space. Zixun Guo is a research student at the UKRI Centre for Doctoral Training in Artificial Intelligence and Music, supported jointly by UK Research and Innovation [Grant number EP/S022694/1] and Queen Mary University of London.

\bibliography{anthology,tango}

\appendix

\clearpage
\onecolumn
\FloatBarrier
\section{Reverse Diffusion Process}
\label{sec:rev_diff}

The reverse process to iteratively reconstruct $z_0$ is as following:  
\begin{flalign}
    &p_\theta^{mus}(z_{n-1}|z_n, \mathcal{C} ) = \mathcal{N}(\mu^{(n)}_\theta(z_n, \mathcal{C} ), \Tilde{\beta}^{(n)}),\label{eq:munet}\\
    &\mu_\theta^{(n)}(z_n, \mathcal{C} ) = \frac{1}{\sqrt{\alpha_n}}[z_n - \frac{1 - \alpha_n}{\sqrt{1 - \overline\alpha_n}}\hat\epsilon_\theta^{(n)}(z_n, \mathcal{C} )],\\
    &\Tilde{\beta}^{(n)} =\frac{1 - \bar{\alpha}_{n-1}}{1 - \bar{\alpha}_n} \beta_n,\\
    &\alpha_n = 1 - \beta_n,\\
    &\overline\alpha_n = \prod_{i=1}^n \alpha_n,\\
    &\hat\epsilon_\theta^{(n)}(z_n, \mathcal{C} ) = w~\epsilon_\theta^{(n)}(z_n, \mathcal{C} ) + (1 - w) \epsilon_\theta^{(n)}(z_n), \label{eq:guidance-scaling}
\end{flalign}
where $w$ is the guidance scale in \cref{eq:guidance-scaling} used during inference. During training however, $\epsilon_\theta^{(n)}(z_n, \mathcal{C} )$ is directly used for noise estimation where the conditions $\mathcal{C} $ are randomly dropped as specified in \cref{sec:trainsetup}.

\section{Training Details} \label{sec:app-training}

To further improve the robustness of the classifier-free guidance in \modelemoji{}, we use these three dropouts during training:
\begin{enumerate}[itemsep=0pt, leftmargin=*, wide, labelwidth=0pt, labelindent=0pt, parsep=0pt]
    \item With 5\% probability, drop all the inputs (text, beats, and chords);
    \item With 5\% probability, drop an input feature (applied to each of the inputs separately);
    \item We determine the probability of masking a prompt as $\min(100, 10 \frac{N}{M})$\%, where $N$ represents the number of sentences in the current prompt, and $M$ is the average number of sentences per prompt. Once a prompt is chosen for masking, we randomly draw an integer $X$ from a uniform distribution in the range [20, 50] and proceed to remove $X$\% of the input sentences in the prompt.
\end{enumerate}
The idea behind the first two dropouts is to enable the model to work with incomplete, faulty, or missing input information. The third dropout is aimed at improving robustness for short text inputs. We apply these dropouts to \texttt{Tango} as well, with a small modification: since \texttt{Tango} does not use music feature inputs, we replace the first two dropouts with a single 10 \% probability of dropping all text.

To train \modelemoji{} and Tango baselines, we used various GPU resources: 4 Nvidia Tesla V100 GPUs, and 8 Quadro RTX 8000 GPUs. Training time ranged from 5 to 10 days with effective batch size of 32.

\section{Performance of the Predictors}
During the inference phase, we utilize pre-trained predictors for chord and beat predictions based on textual prompts. These predictors exhibit exceptional performance when the prompts explicitly contain chord and beat information, achieving accuracy of 94.5 \% on the TestB dataset. However, our interest extends to evaluating their performance in scenarios where control sentences are absent from the prompt—essentially, do these predictors generate noisy chords and beats? The concern is that such noise might propagate from the predictors to \modelemoji{}, significantly impacting the overall quality of the generated music.

In our experiments, TestA serves as a scenario where control sentences are not included in the textual prompts. Upon comparing the performance (\Cref{tab:obj_eval}) of \texttt{Tango} and \modelemoji{} on TestA, we observe that the latter outperforms the former across most metrics. This observation indicates that the control predictors do not compromise the performance of \modelemoji{} relative to \texttt{Tango}. The adaptability of these predictors to specific themes or styles in the absence of control sentences remains a potential avenue for future exploration, a topic we briefly touch upon below.

First, we investigate the effect of the Chord predictor on the generated output in a little comparison experiment. We take both TestA and TestB samples synthesized by \modelemoji{} and extract features from them. Then, we evaluate the chord control metrics of PCM, ECM, CMO, and CMOT using chords predicted by chord predictor vs chords detected in the audio from feature extraction.
The metrics on TestA are PCM - 16.15, ECM - 33.95, CMO - 39.81, and CMOT - 47.82.
The metrics on TestB are PCM - 17.75, ECM - 32.07, CMO - 47.36, and CMOT - 66.80.
These results show that \modelemoji{} tends to follow the chords predicted by the chord predictor quite often. While the results on TestA are a bit lower than on TestB, they are still higher than Tango results on TestB as shown in \Cref{tab:control_eval}.

Second, we take a look at some specific examples:

\begin{mdframed}[backgroundcolor=blue!10] 
\textbf{Prompt:} ``This folk song features a female voice singing the main melody. This is accompanied by a tabla playing the percussion. A guitar strums chords. For most parts of the song, only one chord is played. At the last bar, a different chord is played. This song has minimal instruments. This song has a story-telling mood. This song can be played in a village scene in an Indian movie. \textit{The chord sequence is Bbm, Ab. The beat is 3. The tempo of this song is Allegro. The key of this song is Bb minor.}''

Without control sentences in italics (TestA):
\textbf{chords predicted}: ["G", "C", "G", "C", "G", "C"], \textbf{chords predicted time}: [0.46, 1.21, 3.25, 5.48, 7.24, 8.92].
\textbf{chords extracted from audio}: ["G6", "C", "G", "C", "G", "Cmaj7"], \textbf{chords time extracted from audio}: [0.46, 1.58, 3.07, 5.94, 7.62, 9.66]

With control sentences in italics (TestB):
\textbf{chords predicted}: ["Bbm", "Ab"], \textbf{chords predicted time}: [0.46, 7.24],
\textbf{chords extracted from audio}: ["F\#maj7", "Ab"], \textbf{chords time extracted from audio}: [0.46, 7.43].

\end{mdframed}

\begin{mdframed}[backgroundcolor=blue!10] 
\textbf{Prompt:} ``A female singer sings this bluesy melody. The song is medium tempo with minimal guitar accompaniment and no other instrumentation. The song's medium tempo is very emotional and passionate. The song is a modern pop hit but with poor audio quality. \textit{The key of this song is G minor. The time signature is 3/4. This song goes at 168.0 beats per minute. The chord progression in this song is Am7, G7, Cm, G, A7.}''

Without control sentences in italics (TestA):
\textbf{chords predicted}: ["C\#m7", "C\#m7", "C\#m7", "C\#m7", "C\#m7"], \textbf{chords predicted time}: [0.46, 3.25, 6.32, 8.17, 9.29],
\textbf{chords extracted from audio}: ["F\#", "C\#m", "F\#m", "C\#m7"], \textbf{chords time extracted from audio}: [0.46, 1.21, 4.55, 5.39]

With control sentences in italics (TestB):
\textbf{chords predicted}: ["Am7", "G7", "Cm", "G", "A7"], \textbf{chords predicted time}: [0.46, 1.67, 3.53, 5.48, 8.92],
\textbf{chords extracted from audio}: ["Am", "G", "C", "Gmaj7", "A6", "Gmaj7"], \textbf{chords time extracted from audio}: [0.46, 1.67, 3.72, 5.94, 8.73, 9.85]
\end{mdframed}


The two depicted samples give us some specific insights into the predicted chords and chords detected in the generated audio. Most of the time, \modelemoji{} follows the chords provided by the chord predictor in most cases. We can observe some substitutions in the actual chords detected from the audio compared to the predicted chords, e.g., G became G6, C became Cmaj7, and C\#m7 became C\#m. These chord substitutions are very close musically and could even be a consequence of the feature extraction system not being 100\% accurate. The substitution of Bbm for F\#maj7 is more of a change at first glance, but given that 2 out of 3 notes in Bbm are also contained in the 
4-note F\#maj \modelemoji{}, we see this substitution as understandable too. However, we note that this substitution would not be considered a valid one in any of our proposed chord control metrics.

Last but not least, in the absence of explicit control sentences in the prompt, we observe that the chords predicted by the chord predictor usually follow specific patterns. The generated samples follow a pattern of two chords that alternate (A, B, A, B, A, B). Another type of an observed pattern is one chord repeated (A, A, A, A, A, A). A more elaborate study on the Chord predictor behavior should be a topic for future work.

\section{Insights from the Human Annotation}
Here, we take a look at some generated examples from the expert listening test, specifically a blues sample with the following prompt: \texttt{``An instrumental blues melody played by a lead guitar and a strumming acoustic guitar. The acoustic guitarist's strumming keeps the rhythm steady. The chord sequence is G7, F7, C7, G7. This song goes at 100 beats per minute.''}

\begin{figure}
    \centering
    \includegraphics[width=0.7\textwidth]{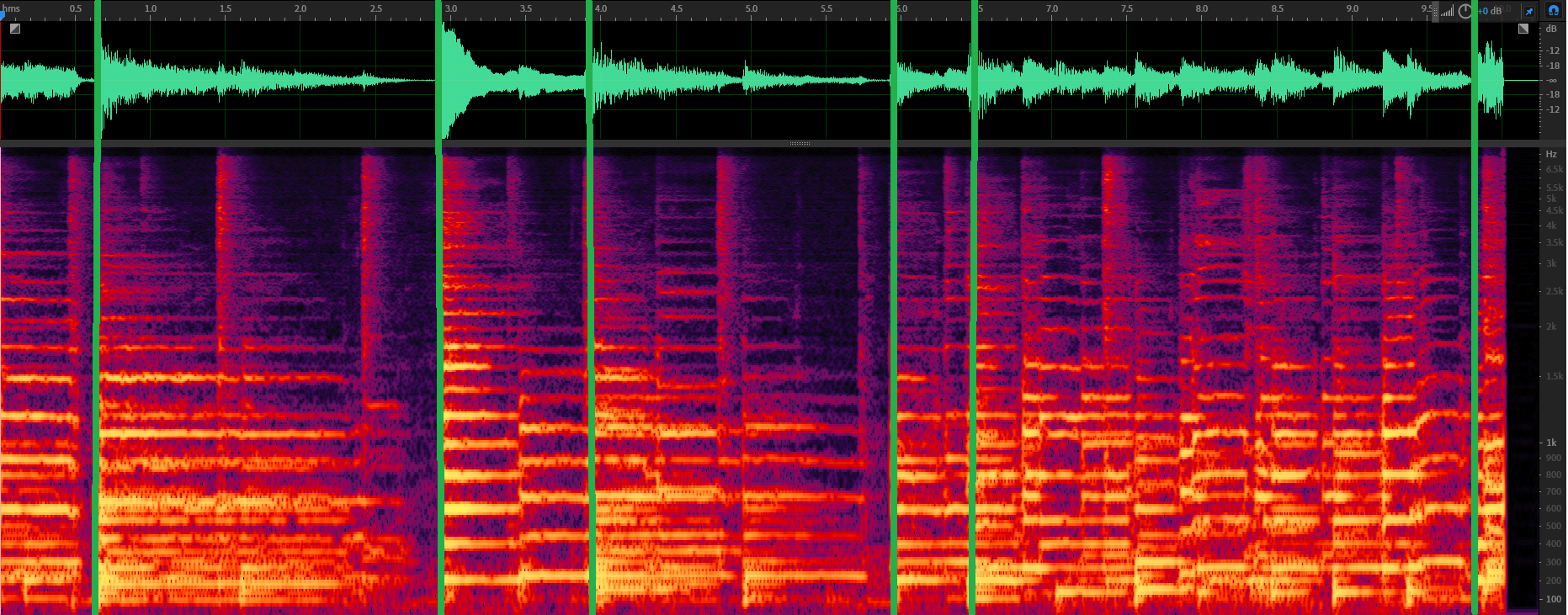}
    \caption{Mel-spectrogram of a blues sample generated by \texttt{Tango} trained on \dataset{}.}
    \label{fig:sample_tango}
\end{figure}

\begin{figure}
    \centering
    \includegraphics[width=0.7\textwidth]{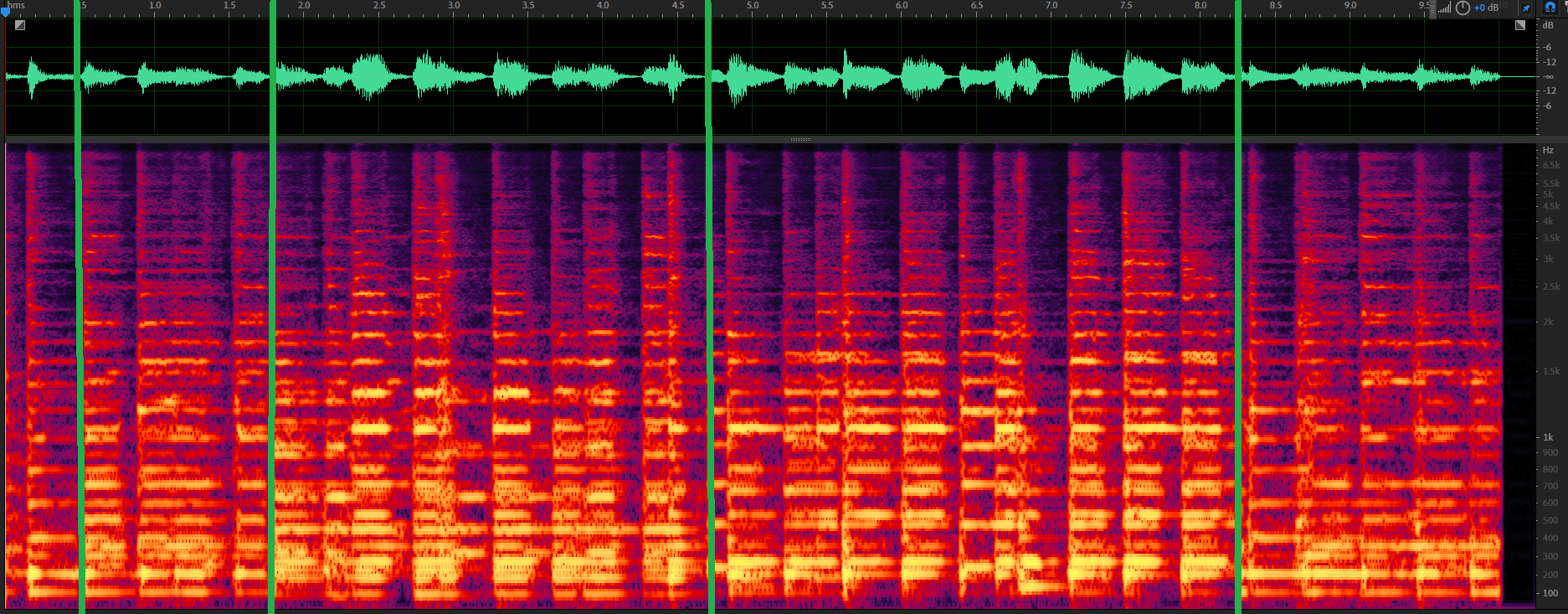}
    \caption{Mel-spectrogram of a blues sample generated by \modelemoji{} with vertical lines showing perceived chord starts.}
    \label{fig:sample_mustango}
\end{figure}

In \Cref{fig:sample_tango} we can see the mel-spectrogram generated by pre-trained \texttt{Tango} finetuned on \dataset{}. As is clear from the spectrogram and the waveform attached, the music appears a bit abruptly in contrast to the sample generated by \modelemoji{} depicted in \Cref{fig:sample_mustango} where the rhythm is very consistent. This seems to reflect the results of our expert listening study from \Cref{tab:subj_eval}. The predicted beat timestamps by our Beat predictor that condition the diffusion process are as follows:
\textbf{beats predicted}: [[0.26, 0.87, 1.52, 2.09, 2.76, 3.41, 4.0, 4.57, 5.1, 5.65, 6.22, 6.79, 7.36, 7.79, 8.3, 8.8, 9.3, 9.75], 3].
\textbf{These predicted beat timestamps show that there is a beat roughly every 0.6 seconds, which corresponds to 100 beats per minute tempo. This is the tempo ordered and properly predicted to condition the model.}

When it comes to chords, Tango would sometimes not follow the chords, make them sound unclear, or not give them enough time to sound through. On the other hand, \modelemoji{} seems to follow the predicted chords as well as their starting time. We take a look at the same blues example. The predicted chord condition from the Chord predictor is as follows:
\textbf{chords predicted}: ["G7", "F7", "C7", "G7"], \textbf{chords predicted time}: [0.46, 2.04, 4.37, 8.17].
We can see that the chord onset time is nicely spread in time. This is also clear from listening to the sample and seeing the spectrogram with perceived chord starts in \Cref{fig:sample_mustango}. To confirm this, we extracted the chord features from the generated audio to compare.
The chord feature extracted from the audio sample generated by \modelemoji{} is:
\textbf{chords}: ["G7", "F7", "C", "G7"], \textbf{chords time}: [0.46, 1.76, 4.74, 8.45]
Interestingly, the match of timing and chord sequence is very clear here. The substitution of the C7 chord for C can be a minor mistake either on the generation part or the feature extraction part. If we consider the chord metrics from the controllability evaluation in \cref{sec:objeval}, this would yield a score of 100 for CMOT and a score of 75 for CMO and ECM.
In contrast, the sample generated by pre-trained Tango finetuned on \dataset{} sounds more unstable and does not give enough time to chords to sound through.
The chord feature extracted from the audio sample generated by pre-trained Tango finetuned on \dataset{} is:
\textbf{chords}: ["Fm6", "G", "Dm", "G", "C", "Gm"], \textbf{chords time}: [0.46, 2.69, 3.53, 5.76, 6.69, 9.66]. We can see that there are 6 chords extracted from the audio sample instead of the ordered 4, and they do not match too well, as we see a minor type of F chord instead of a major; G also appears in a minor variant once; and there is an additional Dm chord too. This would yield a CMOT score of 75, but CMO and ECM scores of 0. The perceived chord starts can be seen in \Cref{fig:sample_tango}.


\section{Related Works}\label{sec:relw}

\vspace{5cm}
\begin{table}
\centering
\begin{adjustbox}{width=\linewidth}
\begin{tabular}{l|p{6cm}|l|p{6cm}}
\toprule
\textbf{Model} & \textbf{Dataset} & \textbf{Core Architecture} & \textbf{Area of Focus} \\
\midrule
MusicLM~\cite{agostinelli2023musiclm} & Private Dataset including an open-sourced test set: MusiCaps & Hierarchical Seq2Seq Modeling & Audio Quality, Text-Music Relevance \\
Noise2Music~\cite{huang2023noise2music} & Private Dataset obtained via pseudo labelling & 2-stage Diffusion & Audio Quality, Text-Music Relevance \\
Ernie-Music~\cite{ERNIE_MUSIC} & Private Dataset consisting of online music and corresponding comments & Diffusion (without using Audio Latent) & Audio Quality, Text-Music Relevance, Diversity \\
MusicGEN~\cite{copet2023simple} & Private Dataset & Autoregressive Transformer & Audio Quality, Text-Music Relevance, Music Quality, Controllability (Follows given melodies) \\
Mousai~\cite{mousai} & Private Dataset; Data Collection Pipeline partially open-sourced & 2-Stage Latent Diffusion & Audio Quality, Text-Music Relevance, Music Quality, Efficiency, Long-Term Structure, Diversity \\
JEN1~\cite{li2023jen} & Private Dataset & Latent Diffusion, Multi-Task Learning & Audio Quality, Text-Music Relevance, Music Quality, Efficiency \\
\bf \model{} (ours) & \textbf{Public Dataset} + Music-Domain-Knowledge-Enhanced Data Augmentation & Latent Diffusion & Audio Quality, Text-Music Relevance, Music Quality, \textbf{Music Controllability} (Follows user-specific text prompts including tempo, chord changes, etc) \\
\bottomrule
\end{tabular}
\end{adjustbox}
\caption{High-level comparison among various recent text-to-music models.}
\label{tab:model_comp}
\end{table}

\vspace{-5cm}
In this section, we describe existing state-of-the-art research on text-to-audio generation, followed by the more specific domain of text-to-music generation.
For audio generation, AudioLM~\cite{borsos2023audiolm} uses the state-of-the-art semantic model w2v-Bert~\cite{DBLP:conf/asru/ChungZHCQPW21} to generate the semantic tokens from audio prompt. These tokens condition the generation of acoustic tokens that are decoded using acoustic model SoundStream~\cite{DBLP:journals/taslp/ZeghidourLOST22} to generate audio.

AudioLDM~\cite{liu2023audioldm} is a text-to-audio framework that leverages CLAP~\cite{wu2023large}, a joint audio-text representation model, and a latent diffusion model (LDM). Specifically, an LDM is trained to generate the latent representations of melspectrograms which are obtained using a VAE. During diffusion, the CLAP embeddings are utilized to guide the generation. \texttt{Tango} \cite{ghosal2023tango} leverages the pre-trained VAE from AudioLDM and replaces the CLAP model with an instruction fine-tuned large language model: FLAN-T5 to achieve comparable or better results while training with a much smaller dataset.

In the field of music generation, there is a long history of generated MIDI music \cite{herremans2017functional}. Using MIDI may be useful for producers to work with in Digital Audio Workstations, yet it has the disadvantage that datasets are extremely limited. In recent years, the focus of conditional music generation within the audio domain has centered around musical conditions, such as note intensity or tempo~\cite{Musika}. More recently, however, models that directly generate \textit{audio} music from text captions have emerged. A summary of these papers are provided in Table~\ref{tab:model_comp}. MusicLM~\cite{agostinelli2023musiclm} uses two pre-trained models, MuLan~\cite{DBLP:conf/ismir/HuangJLGLE22}, a joint text-music embedding model, and w2v-Bert~\cite{DBLP:conf/asru/ChungZHCQPW21}, a masked language model to address the challenge of maintaining both synthesizing quality and coherence during music generation. These two pre-trained models are then utilized to condition the acoustic model SoundStream~\cite{DBLP:journals/taslp/ZeghidourLOST22} which in turn can generate acoustic tokens autoregressively. These acoustic tokens are then decoded by SoundStream to become the final audio output. MusicLM outperforms two existing commercially available text-to-music software: Mubert\footnote{\url{https://github.com/MubertAI/Mubert-Text-to-Music}} and Riffusion\footnote{\url{https://www.riffusion.com/}} in terms of Frechet Audio Distance, Faithfulness to the text description, KL divergence, and Mulan Cycle Consistency. Since no publications are linked to these latter two systems, the model details are not available.

Another text-to-music model is Noise2Music~\cite{huang2023noise2music}. To obtain training data for the model, the authors propose a method to obtain a large amount of paired music and text data in which LaMDA-LF~\cite{DBLP:journals/corr/abs-2201-08239}, a large language model, is used to generate multiple generic candidate text descriptions. The aforementioned joint text-music embedding MuLan is then utilized to select the best candidates for existing music data. The obtained music and text pairs are then used to train a two-stage diffusion model, where the first diffusion model generates an intermediate representation and the second generates the final audio output. 

Ernie-Music~\cite{ERNIE_MUSIC} uses a diffusion model to generate music audio from free-form text. It is trained using a private dataset which consists of online music and the top-rated comments from the comment section. The authors recruited 10 casual music listeners to participate in a listening study. Results showed that Ernie-Music outperforms two non-diffusion-based generative systems. 



In recent months, a number of text-to-music models have come out. \citet{mousai} proposes a 2-stage diffusion model in which the first diffusion magnitude autoencoder (DMAE) learns a meaningful latent representation of music (64 times smaller than the input), while in the second diffusion model, text condition along with the latent acquired at the first stage is included to guide the final music generation. MusicGen~\cite{copet2023simple} utilizes a single-stage transformer LM with efficient token interleaving patterns to achieve high-quality generation and better controlabillity over the output. MusicGen can be conditioned by a text prompt, or by an audio fragment in the form of a chromagram. The system was trained with a licensed dataset. The JEN-1 model \cite{li2023jen} is an omnidirectional diffusion model designed to perform various tasks such as text-guided music generation, music inpainting, and continuation. 
Another interesting recent model is that of \citet{su2023v2meow}, which focuses on generating music pieces to complement video, conditioned on both video and text inputs. Unlike text, video conditioning can contain a lot of temporal information, such as beats and emotions, which are important for music.

\section{FMACaps dataset creation}\label{sec:fmacaps}
We source the new music files from the Free Music Archive (FMA) \cite{defferrard2016fma}, a large dataset of popular songs. In particular, we took 1,000 random samples from FMA-large and clipped out a random 10-second fragment from each of them. Then, we used Essentia's tagging models~\cite{bogdanov2013essentia} to assign tags to audio. Specifically, we used the models for general auto-tagging, mood, genre, instrumentation, voice, and voice gender which provide us with a rich set of tags along with their probabilities. Then, a music expert wrote text descriptions for 25 of the samples based on the audio as well as the extracted tags. Next, we instructed ChatGPT 
to perform an in-context learning task to get pseudo-prompts from tags for the rest of the dataset. Finally, we added relevant control sentences to the prompts after extracting relevant music features, as described in \cref{sec:method_feature_extract}. Similar to our training set, we added 0/1/2/3/4 control sentences with a probability of 25/30/20/15/10\% respectively. We refer to this evaluation set as \datasetFMA{}.

\section{User Interface and Questions used for Listening Studies}\label{sec:eval-interface}
\FloatBarrier

\begin{figure}[h]
    \centering
    \includegraphics[width=0.65\textwidth]{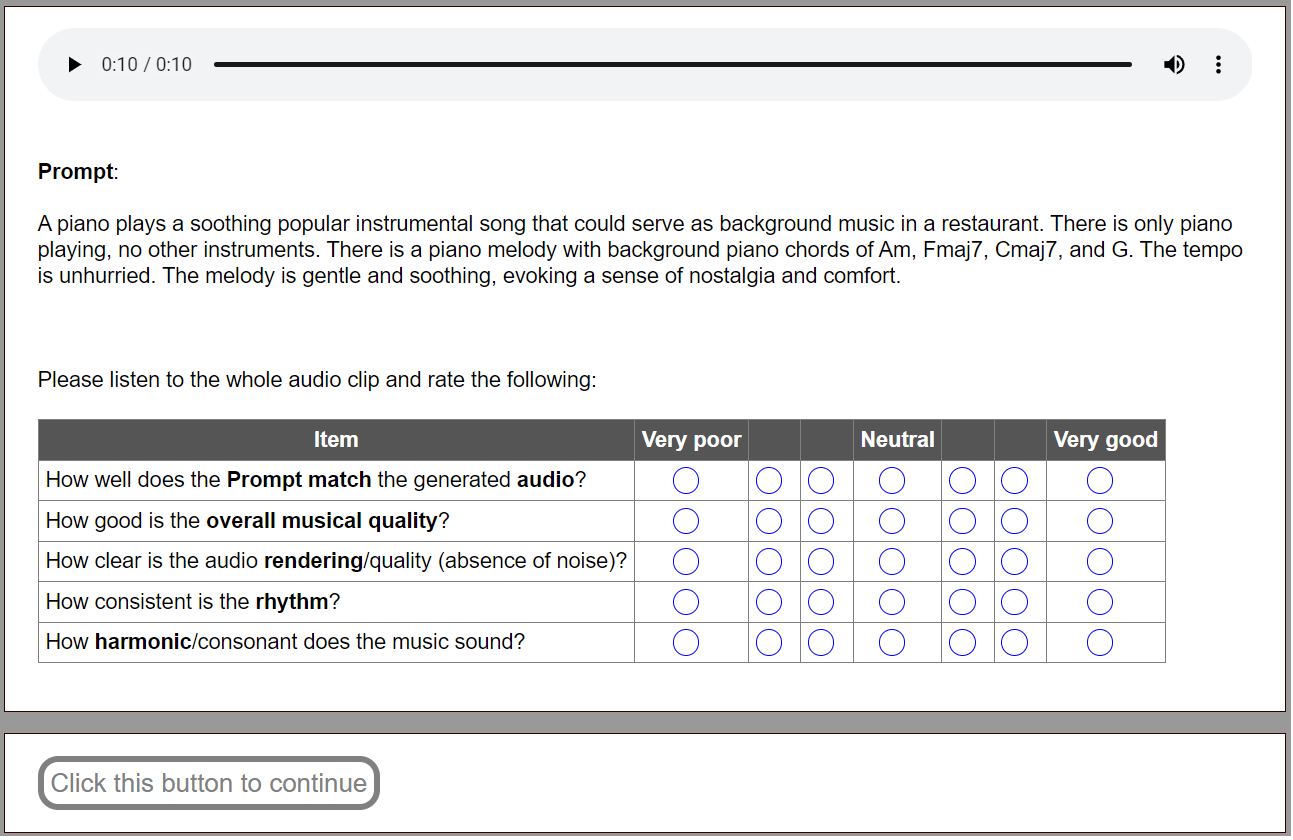}
    \caption{Question interface used for the general listening test. }
    \label{appfig:general_study}
\end{figure}

\begin{figure}[h]
    \centering
    \includegraphics[width=0.65\textwidth]{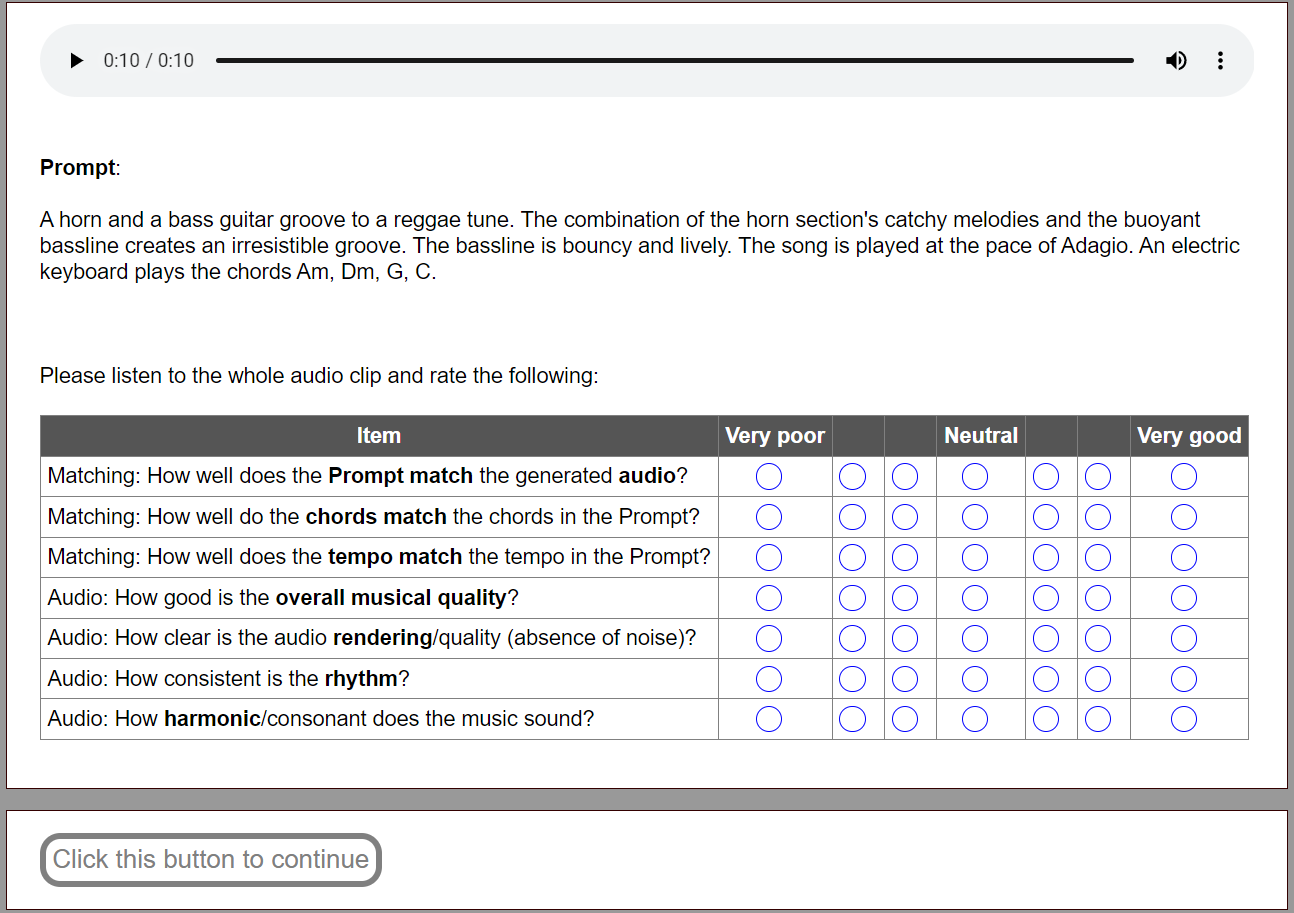}
    \caption{Question interface used for the controllability listening test with music experts. }
    \label{appfig:expert_study}
\end{figure}

The human evaluation participants for the listening tests were recruited through email contacts of various music research and machine learning communities. Their age ranged from 15 to 60+ years. No specific information on the country of residence was collected. Participation was free of reward and voluntary. All the participants were informed about this before their participation. The data collection protocol for the listening test was approved by an independent ethics review board.

\clearpage
\FloatBarrier
\section{ChatGPT Rephrasing}\label{sec:app-chatgpt}
In our data augmentation pipeline, we rephrase captions using ChatGPT. The instructions to the API are as follows:

\begin{mdframed}[backgroundcolor=blue!10] 
"I have a song for which the caption is the following:"
*caption*
"I have made some changes to the audio file which are optionally described towards the end of the caption. Can you rephrase the caption more naturally in a single paragraph using all the musical terms provided above? You should generate only the caption and nothing else. Do not use the word modification in your generation. The length of the new caption should be no more than eight sentences."
\end{mdframed}

\vspace{5mm}

\textbf{Examples of rephrasing:}
\begin{mdframed}[backgroundcolor=yellow!10] 
\textbf{Original}

This folk song features a male voice singing the main melody in an emotional mood. This is accompanied by an accordion playing fills in the background. A violin plays a droning melody. There is no percussion in this song. This song can be played at a Central Asian classical concert. The chord sequence is B, Dmaj7, B, Gbm. This song goes at 95.0 beats per minute. The beat counts to 2. This song is in the key of B minor.
\end{mdframed}

\begin{mdframed}[backgroundcolor=green!10] 
\textbf{Rephrased}

A male voice sings the main melody with emotional intensity in this Central Asian folk song. The accordion fills the background while a violin provides a droning melody. There is no percussion in this piece, which can be played at a classical concert in Central Asia. With a chord sequence of B, Dmaj7, B, Gbm, the song has a beat count of 2 at a tempo of 95.0 beats per minute. The song is in the key of B minor.
\end{mdframed}

\begin{mdframed}[backgroundcolor=yellow!10] 
\textbf{Original}

This folk song features a male voice singing the main melody in an emotional mood. This is accompanied by an accordion playing fills in the background. A violin plays a droning melody. There is no percussion in this song. This song can be played at a Central Asian classical concert. The key is C minor. The chord progression in this song is C, D\#maj7, C, Gm.
\end{mdframed}

\begin{mdframed}[backgroundcolor=green!10] 
\textbf{Rephrased}

This emotional folk song, perfect for a Central Asian classical concert, showcases a male voice singing the main melody accompanied by a droning violin and accordion fills in the background. With no percussion present, the key of C minor sets the tone, and the chord progression follows suit with C, D\#maj7, C, Gm.
\end{mdframed}


\FloatBarrier
\section{Control-Sentence Templates to Enhance the Prompts}
\label{sec:controlcaptions}

\begin{table}[h] \footnotesize
\begin{tabular}{@{}lll@{}}
\toprule
Feature & Input   & Output sentences  \\
\midrule
Tempo & int $i$ & \tabitem The bpm is $i$. \\
& & \tabitem The tempo of this song is $i$ beats per minute.\\
& & \tabitem This song goes at $i$ beats per minute.\\
\midrule
Tempo & string $w \in $ [`Grave', `Largo', `Adagio', & \tabitem This song is in $w$.\\
& `Andante', `Moderato', `Allegro',  & \tabitem The tempo of this song is $w$.\\
& `Vivace', `Presto', `Prestissimo']& \tabitem This song is played in $w$.\\
& & \tabitem The song is played at the pace of $w$.\\
\midrule
Beat count& int $b$ & \tabitem The time signature is $\frac{b}{4}$.\\
& & \tabitem The beat is $b$.\\
& & \tabitem The beat counts to $b$.\\
\midrule
Chords&  text list of chords $s$ & \tabitem The chord sequence is $s$. \\
& & \tabitem The chord progression in this song is $s$.\\
\midrule
Key& string $rootnote$  & \tabitem The key is $rootnote$ $m$\\
& string $m \in$ [`major', `minor'] & \tabitem The key of this song is $rootnote$ $m$.\\
& & \tabitem This song is in the key of $rootnote$ $m$\\
\midrule
Volume change& float $f$ indicating start/end time  & \tabitem There is a $w$ from start until $f$ seconds \\
& of crescendo/decrescendo, & \tabitem The song starts with a $w$.\\
& string $w$ $\in$ ['crescendo', 'decrescendo'], & \tabitem $u$ the volume progressively!\\
& and $u \in $ ['increase', 'decrease'] & \tabitem There is a $w$ from $f$ seconds on.\\
& & \tabitem At seconds $f$, the song starts to gradually $u$ in volume.\\
& & \tabitem Midway through the song, a $w$ starts.\\
\bottomrule
\end{tabular}
\caption{Rules used to create text sentences from input parameters detected from the data (key, chords, beats, tempo), and those used to augment the data (crescendo, etc.). Note that the tempo strings $w$ were assigned based on music-theory binning in terms of bpm: Grave (0, 40], Largo (40, 60], Adagio (60, 70], Andante (70, 90], 
Moderato (90, 110], 
Allegro (110, 140], 
Vivace (140, 160], 
Presto (160, 210],
Prestissimo (210, $\infty$).}
\label{tab:app_attr_to_text}

. 
\end{table}


\clearpage
\FloatBarrier
\section{Custom Captions used for Listening Studies}
\label{sec:captions}
\FloatBarrier

\begin{table*}[h]
\footnotesize
\begin{tabular}{p{0.4cm} p{15cm}} 
\toprule
1& This piece is an instrumental reggae song that is very chill and slow. There is no singer. It is relaxing to hear the groove with the bass guitar. The song includes reggae electric guitar, horn, and percussion like bongos. The keyboard provides lush chords. The time signature is 4/4. The chord progression is G, F, C.\\ 
2& This instrumental blues song goes very slow at a bpm of 50. You can hear the bass, harmonica and guitar grooving. The harmonica plays a solo over the harmonious guitar and bass.  \\
3 & This classical piece is a waltz played by a string quartet. It includes two violins, a viola, and a cello, the beat counts to 3. It sounds elegant, and has a strong first beat. It has a natural and danceable rhythm. The mood is romantic. The chord progression is Em, Am, D, G. \\
4& African drums are playing a complex rhythm while a male vocalist chants a ritual. The atmosphere is mesmerizing. The complex drumming pattern is a mesmerizing blend of syncopation, polyrhythms, and intricate patterns. It takes place somewhere in the wilderness, or in an indigenous village. \\
5& This rock piece with guitars and drums is loud but fades out later on and becomes softer. It sounds powerful yet melancholic. It is instrumental only. A bass guitar provides a steady beat, enhancing the groove and energy of the song.    \\
6& A single bass instrument is playing a running baseline. It has a jazzy feeling to it and sounds mellow. This could be played in a jazz club. The tempo is 120 bpm.    \\
7& This is a hip hop song. It has two rappers taking turns, one female and one male. An electronic synth melody sample in the background keeps on looping. We can hear electronic beats and sometimes record-scratching sound effects.     \\
8& A smooth jazz song with saxophone, drums and guitar with a chord progression of Dm7, G7, Cmaj7. The song is relaxed and slow. There are no vocals, it is instrumental only. The saxophone produces a velvety tone that delivers an emotive melody. \\
9& A piano plays a soothing popular instrumental song that could serve as background music in a restaurant. There is only piano playing, no other instruments. There is a piano melody with background piano chords of Am, Fmaj7, Cmaj7, and G. The tempo is unhurried. The melody is gentle and soothing, evoking a sense of nostalgia and comfort.    \\
10& Indian folk music with a sitar and female vocals. It evokes a sense of zen and elevation. A sitar player begins with a gentle and melodic introduction, plucking the strings with precision and emotion. There are rhythmic beats of traditional hand percussion instruments, such as the tabla. It could be played at a cultural festival to showcase Indian culture.    \\
\midrule
11& This is a melodic and energetic rock ballad with a male vocalist. It has a country vibe and is of alternative or popfolk genre. The electric and acoustic guitars and the bass create the background, while the drums give a regular beat. The singer's voice is complemented by a piano. \\
12& This is a slow classical piece with violins and pianos. It has a film score feel and is instrumental only. The orchestration is soft, with strings and flutes. \\
13& This fast and energetic rock song is performed by a male singer. The genre is alternative or punk rock. The background is formed by a guitar, an electric guitar, bass, and drums. There is also a synthesizer. \\
14& This is a slow and ambient instrumental piece with a soundscape that feels like space. The atmosphere is meditative and relaxing but with a certain darkness to it. The genre is electronic soundtrack, and the music is completely instrumental with a synthesizer, bass, and drums forming the background. This song goes at 167.0 beats per minute. \\
15& This is an instrumental piece with Indian and classical elements. The sitar, violin, and flute play prominent roles in creating a meditative and relaxing mood. The percussion and guitar provide a background rhythm to this world and jazz fusion. \\ \\
\bottomrule
\end{tabular}
\caption{Custom captions used for the general listening test. Captions in the top part were used in both first and second runs, captions in the bottom part were used in the second run only.}
\end{table*}

\begin{table*}[h]
\footnotesize
\begin{tabular}{p{0.4cm} p{15cm}} 
\toprule
1 & An instrumental blues melody played by a lead guitar and a strumming acoustic guitar. The acoustic guitarist's strumming keeps the rhythm steady. The chord sequence is G7, F7, C7, G7. This song goes at 100 beats per minute.       \\ 
2 & An instrumental blues melody played by a lead guitar and a strumming acoustic guitar. The acoustic guitarist's strumming keeps the rhythm steady. The chord sequence is Dm, Am, Em. This song goes at 60 beats per minute.                   \\
3& A piano plays a popular melody over the chords of Am, Fmaj7, Cmaj7, G. There is only piano playing, no other instruments or voice. The tempo is Adagio.    \\
4& A piano plays a popular melody over the chords of Gm, Bb, Eb. There is only piano playing, no other instruments or voice. The tempo is Vivace.                                                                                                        \\
5& This is an intense and loud punk song with guitars and drums. It is instrumental only. It is very energetic and powerful. The thunderous beats of the drummer provide a pounding rhythm. A guitar solo melody emerges from the chaotic background of the chords. The chord progression is A, D, E. The tempo of the song is 160 bpm.           \\
6& This is an intense and loud punk song with guitars and drums. It is instrumental only. It is very energetic and powerful. The thunderous beats of the drummer provide a pounding rhythm. A guitar solo melody emerges from the chaotic background of the chords.The chord progression is C, B, A, G. The tempo of the song is 100 bpm.          \\
7& A slow paced jazz song played by a saxophone, piano, guitar and drums follows a chord progression of Em7b5, A7, Dm7. The pianist produces delicate harmonies and subtle embellishments. The drummer provides a brushed rhythm. The guitar strums softly, while the saxophone plays a solo over the chords. This song goes at 80 beats per minute.  \\
8& A slow paced jazz song played by a saxophone, piano, guitar and drums follows a chord progression of B7, G7, E7, C7. The drummer provides a brushed rhythm. The guitar strums softly, while the saxophone plays a solo over the chords. This song goes at 115 beats per minute.  \\
9& This is a techno piece with drums and beats and a leading melody. A synth plays chords. The music kicks off with a powerful and relentless drumbeat. Over the pounding beats, a leading melody emerges. It has strong danceability and can be played in a club. The tempo is 120 bpm. The chords played by the synth are Am, Cm, Dm, Gm.    \\
10& This is a techno piece with drums and beats and a leading melody. A synth plays chords. The music kicks off with a powerful and relentless drumbeat. Over the pounding beats, a leading melody emerges. It has strong danceability and can be played in a club. The tempo is 160 bpm. The chords played by the synth are C, F, G.          \\
11& A horn and a bass guitar groove to a reggae tune. The combination of the horn section's catchy melodies and the buoyant bassline creates an irresistible groove. The bassline is bouncy and lively. The song is played at the pace of Adagio. An electric keyboard plays the chords Am, Dm, G, C.                    \\
12& A horn and a bass guitar groove to a reggae tune. The combination of the horn section's catchy melodies and the buoyant bassline creates an irresistible groove. The bassline is bouncy and lively. The song is played at the pace of Moderato. An electric keyboard plays the chords E, B, A.                                                \\
13& This is a metal song with a guitar, drums and bass guitar. The bassist, wielding a solid-bodied bass guitar, adds depth and power to the sonic landscape. The drummer commands a massive drum kit. With a relentless force, they pound out thunderous rhythms, driving the music forward. As the song begins, the guitar roars to life, delivering a series of distorted chords. It follows the chords of Em, C, G, D. The tempo is 120 bpm.   \\
14& This is a metal song with a guitar, drums and bass guitar. The bassist, wielding a solid-bodied bass guitar, adds depth and power to the sonic landscape. The drummer commands a massive drum kit. With a relentless force, they pound out thunderous rhythms, driving the music forward. As the song begins, the guitar roars to life, delivering a series of distorted chords. It follows the chords of A, F\#m, D, E. The tempo is 170 bpm.  \\
15& A man sings a captivating folk song while strumming chords on an acoustic guitar. This fits a campfire evening happening. The chord progression is G, C, D, G. The tempo is 100 beats per minute.                                          \\
16& A man sings a captivating folk song while strumming chords on an acoustic guitar. This fits a campfire evening happening. The chord progression is Am, Em, Dm, Am. The tempo is 70 beats per minute.                                    \\
17& This is a classical music piece played by a string trio. The instruments involved are violin, viola, and cello. The violin plays the lead melody. The cello's soulful and melodic contributions add depth and gravitas to the performance. The time signature is ¾. The tempo of this song is Presto. The chord sequence is E, C\#m, A, B.           \\
18& This is a classical music piece played by a string trio. The instruments involved are violin, viola, and cello. The violin plays the lead melody. The cello's soulful and melodic contributions add depth and gravitas to the performance. The time signature is 4/4. The tempo of this song is Andante. The chord sequence is Am, Dm, E7, Am.   \\
19& This is a pop song with a female singer singing the leading melody and synthesizers looping samples as background. These loops provide the song's electronic foundation, creating a rich and layered sonic landscape. The charismatic female singer has a dynamic and emotive voice. The tempo is Moderato. The chord sequence is                                                                                                               C, G, Am, F. \\
20& This is a pop song with a female singer singing the leading melody and synthesizers looping samples as background. These loops provide the song's electronic foundation, creating a rich and layered sonic landscape. The charismatic female singer has a dynamic and emotive voice. The tempo is Presto. The key is A minor and the chord sequences are Am, Dm, E.   \\ \bottomrule
\end{tabular}
\caption{Custom opposing captions created for the control experiment. }
\label{tab:controlprompts}
\end{table*}

\cref{tab:controlprompts} presents the 20 text prompts used for the expert listening studies. They consist of 10 contrasting pairs written by music experts. Care was given to make sure that they were realistic and that there were no contradicting elements in the prompts. For instance, caption 1 in \cref{tab:controlprompts} contrasts with caption 2. They share the same original caption \textit{``An instrumental blues melody played by a lead guitar and a strumming acoustic guitar. The acoustic guitarist’s strumming keeps the rhythm steady.''}. However, the control sentences are different: {\it ``The chord sequence is G7, F7, C7, G7. This song goes at 100 beats per minute.''} versus {\it ``The chord sequence is Dm, Am, Em. This song goes at 60 beats per minute.''}. Both chord sequences come from blues progressions, but they belong to a different key/mode. The tempo of caption 2 is significantly slower. Such captions are ideally suited to test if the control sentences influence the generated music.

\clearpage
\FloatBarrier
\section{Additional Examples of Generated Music }
\label{app:examples}
\FloatBarrier
Here we show additional samples generated from pre-trained \texttt{Tango} fine-tuned on MusicCaps, \texttt{Tango} finetuned on \dataset{}, \modelemoji{}, \texttt{MusicGen-M}, and \texttt{AudioLDM2}, all generated from the same prompts.\\

\textbf{Prompt:} A horn and a bass guitar groove to a reggae tune. The combination of the horn section's catchy melodies and the buoyant bassline creates an irresistible groove. The bassline is bouncy and lively. The song is played at the pace of Adagio. An electric keyboard plays the chords Am, Dm, G, and C.


\begin{figure}[h]
    \centering
    \includegraphics[width=0.7\textwidth]{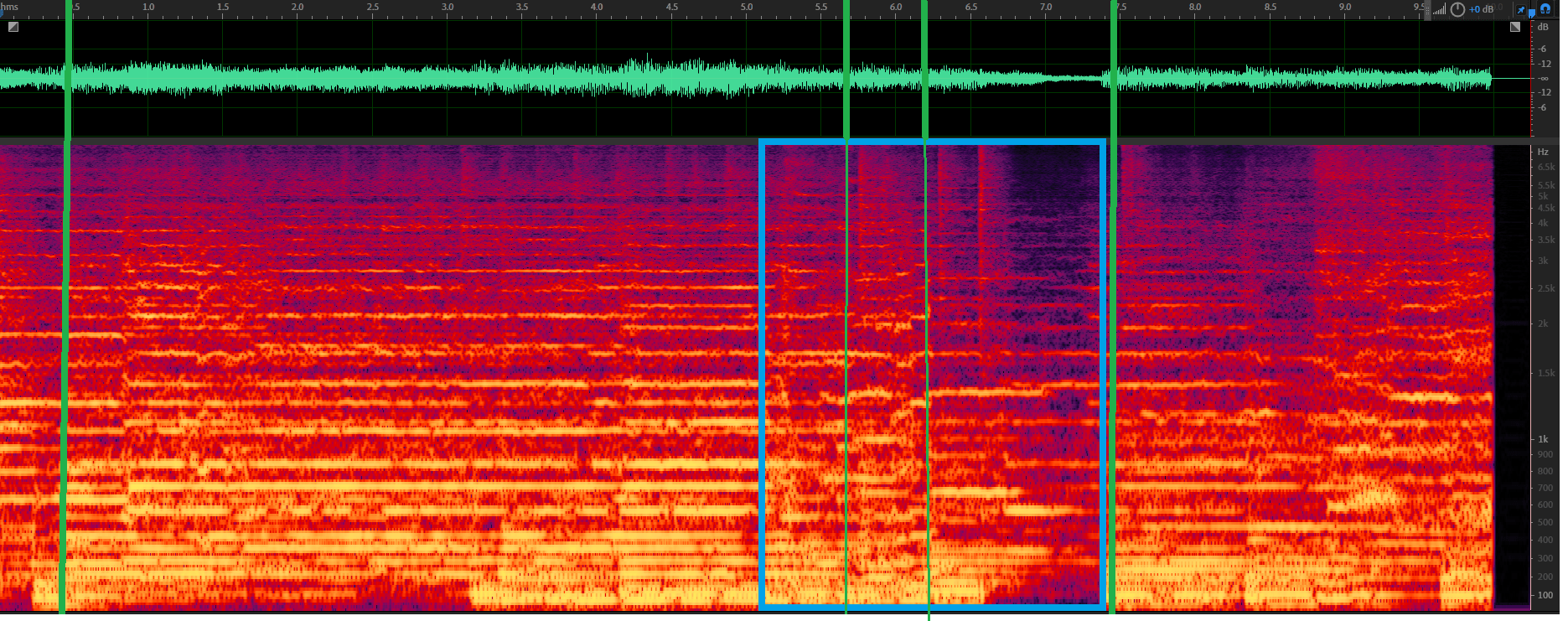}
    \caption{Mel-spectrogram of a reggae sample generated by pre-trained \texttt{Tango} fine-tuned on MusicCaps with vertical lines showing perceived chord starts. The blue box shows an area of dissonance in the music. Overall, the audio is a bit noisy.}
    \label{fig:sample_tango1}
\end{figure}

\begin{figure}[h]
    \centering
    \includegraphics[width=0.7\textwidth]{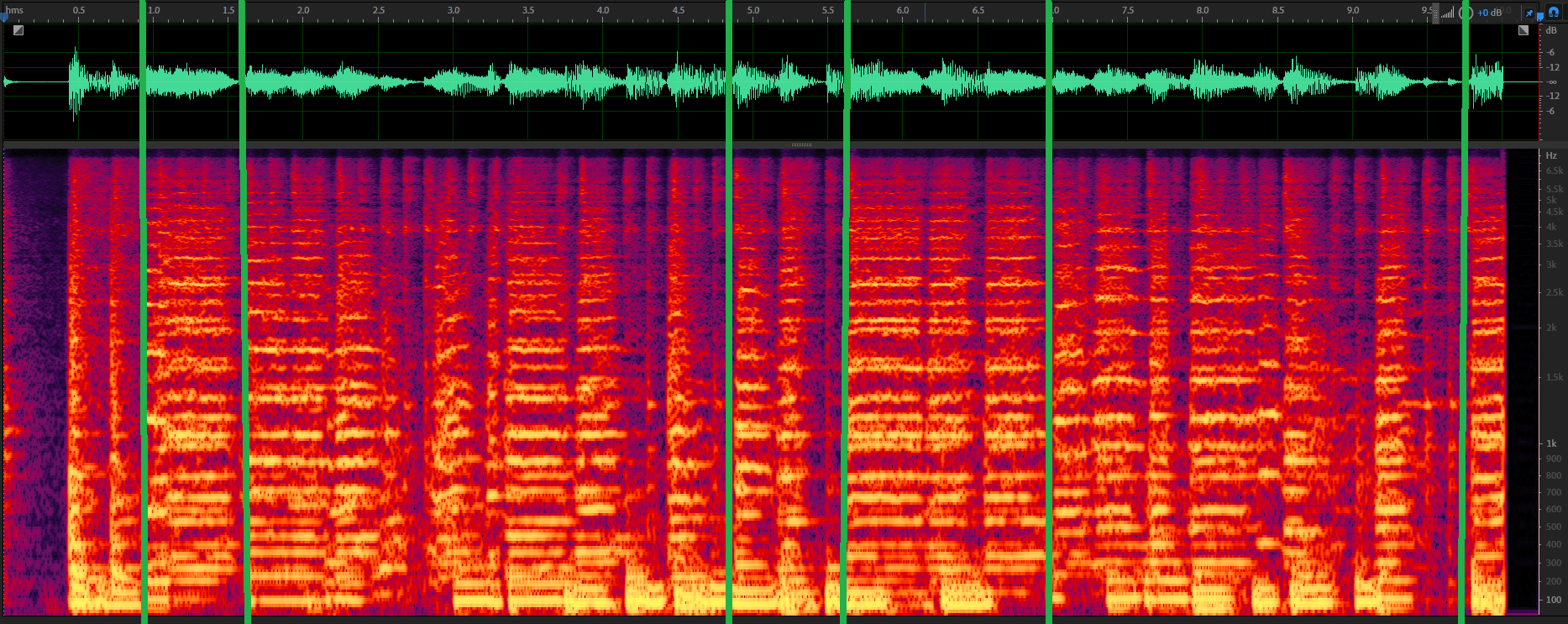}
    \caption{Mel-spectrogram of a reggae sample generated by pre-trained \texttt{Tango} fine-tuned on \dataset{} with vertical lines showing perceived chord starts. There are too many chords here.}
    \label{fig:sample_tangoaug1}
\end{figure}

\begin{figure}[h]
    \centering
    \includegraphics[width=0.7\textwidth]{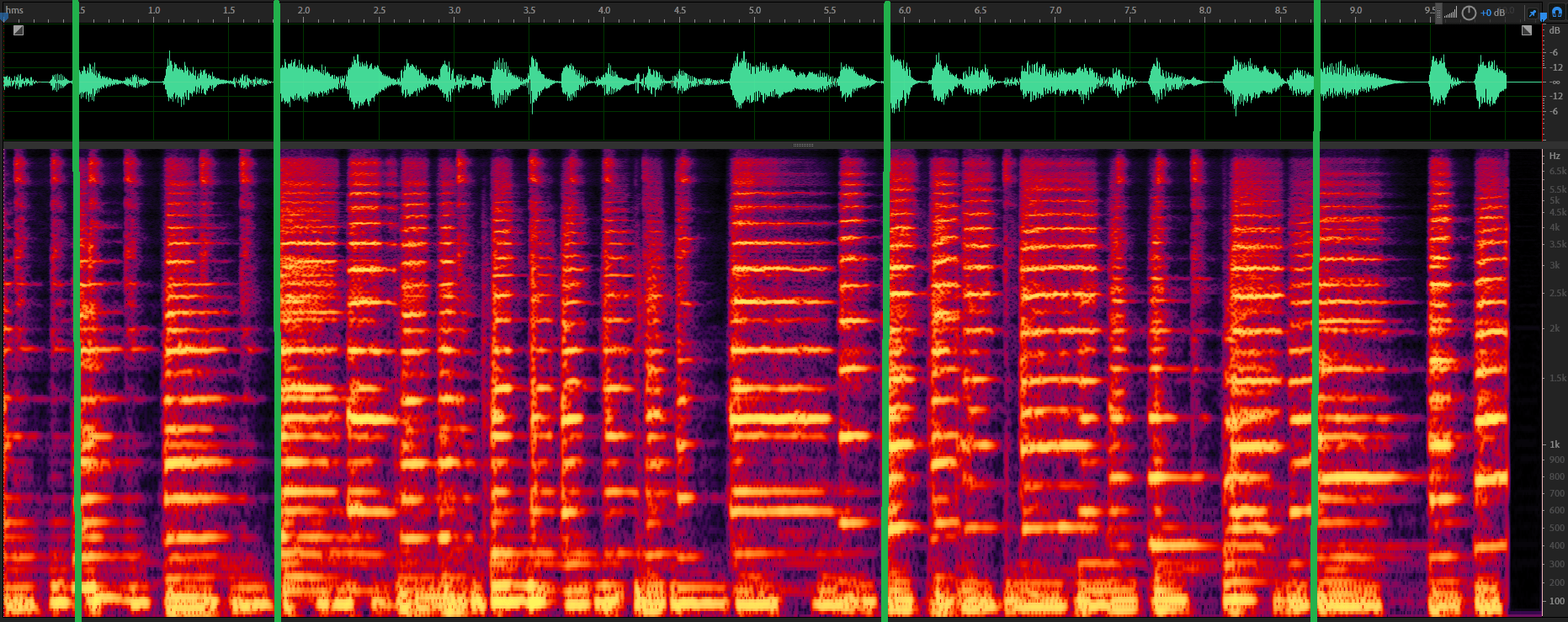}
    \caption{Mel-spectrogram of a reggae sample generated by \model{} with vertical lines showing perceived chord starts. The chords match the prompt. }
    \label{fig:sample_mustango1}
\end{figure}

\clearpage
\FloatBarrier

\textbf{Prompt:} This is a metal song with a guitar, drums and bass guitar. The bassist, wielding a solid-bodied bass guitar, adds depth and power to the sonic landscape. The drummer commands a massive drum kit. With a relentless force, they pound out thunderous rhythms, driving the music forward. As the song begins, the guitar roars to life, delivering a series of distorted chords. It follows the chords of Em, C, G, D. The tempo is 120 bpm.


\begin{figure}[h]
    \centering
    \includegraphics[width=0.7\textwidth]{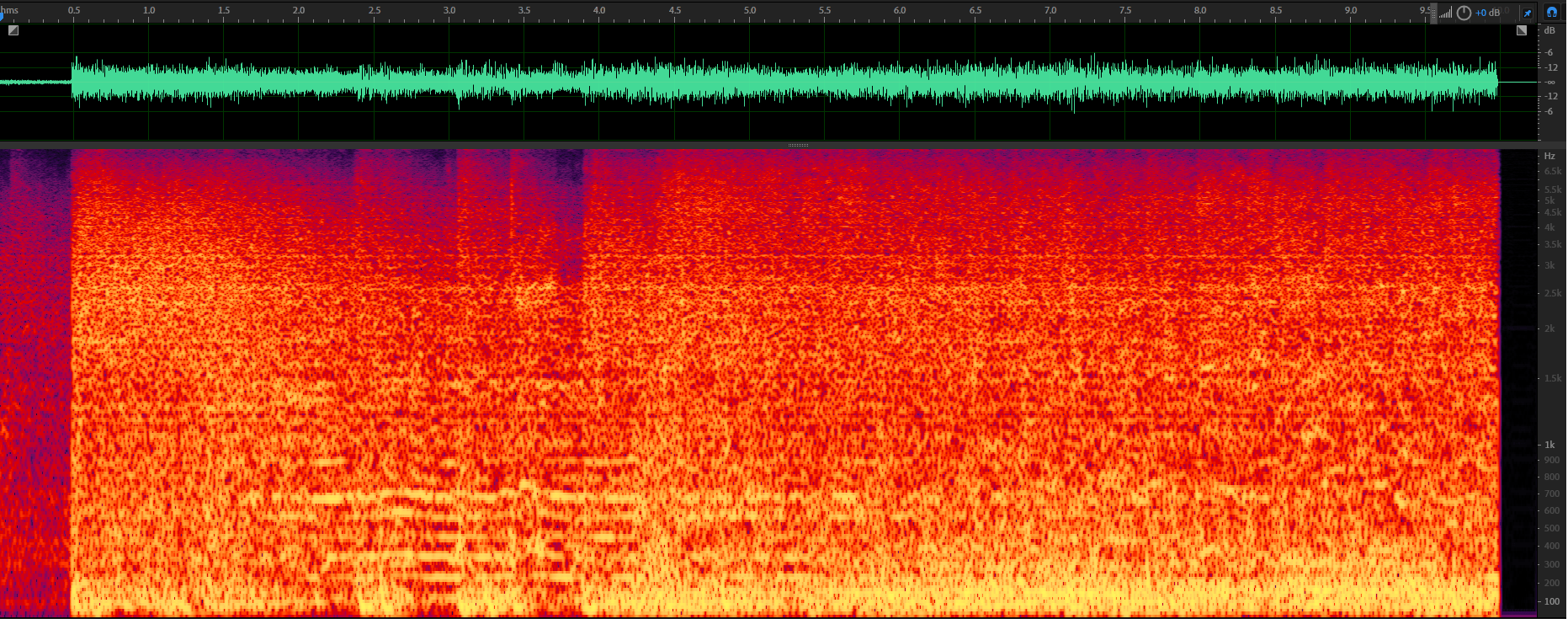}
    \caption{Mel-spectrogram of a metal song sample generated by pre-trained \texttt{Tango} fine-tuned on MusicCaps. It is very noisy from the very start.}
    \label{fig:sample_tango2}
\end{figure}

\begin{figure}[h]
    \centering
    \includegraphics[width=0.7\textwidth]{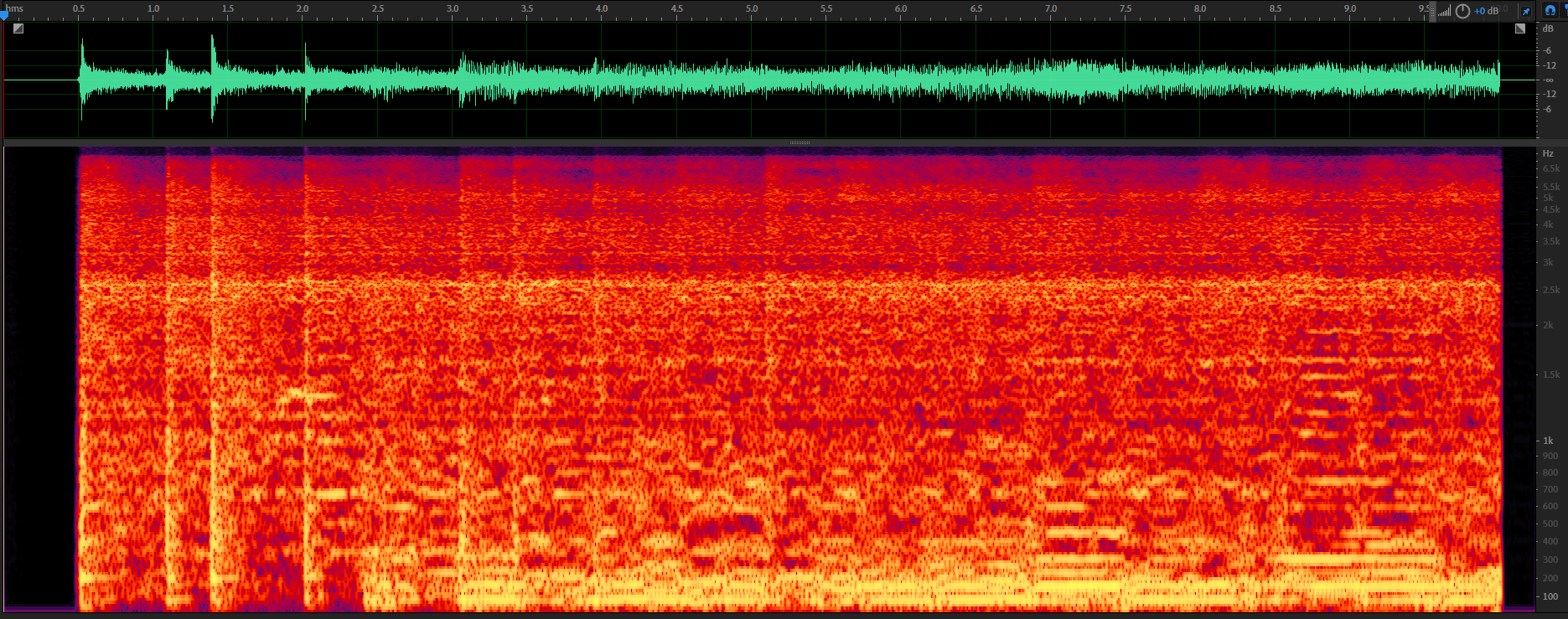}
    \caption{Mel-spectrogram of a metal song sample generated by pre-trained \texttt{Tango} fine-tuned on \dataset{}. The song starts with 4 beats from the drummer, but there is a bit of noise from the start.}
    \label{fig:sample_tangoaug2}
\end{figure}

\begin{figure}[h]
    \centering
    \includegraphics[width=0.7\textwidth]{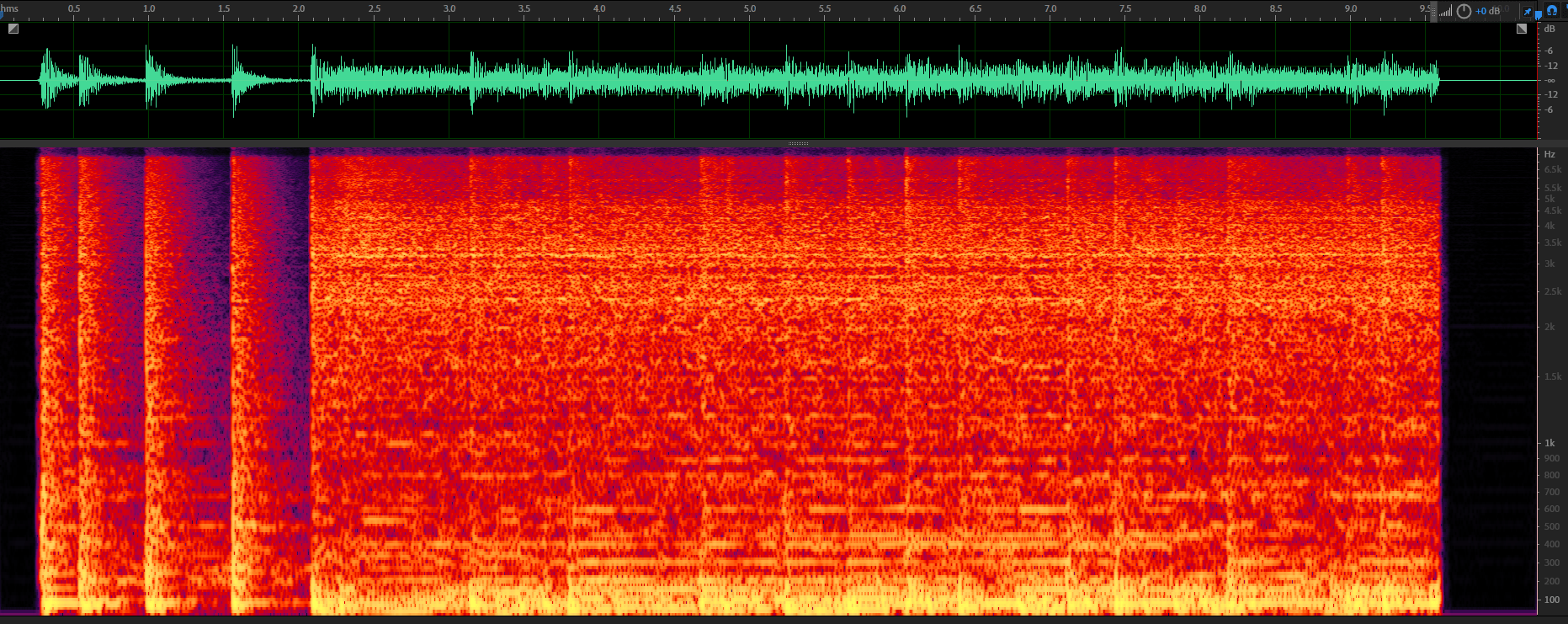}
    \caption{Mel-spectrogram of a metal sample generated by \model{}. The song starts with 4 distinguishable beats from the drummer, then the guitars join.}
    \label{fig:sample_mustango2}
\end{figure}

\clearpage
\FloatBarrier


\textbf{Prompt:} This is a classical music piece. There are violins playing a lead theme, with a double bass and cymbals in the background. It is a melancholic, rather sad piece. The music builds up in volume gradually. The key is A minor. The chord sequence is Am, C, Am.


\begin{figure}[h]
    \centering
    \includegraphics[width=0.7\textwidth]{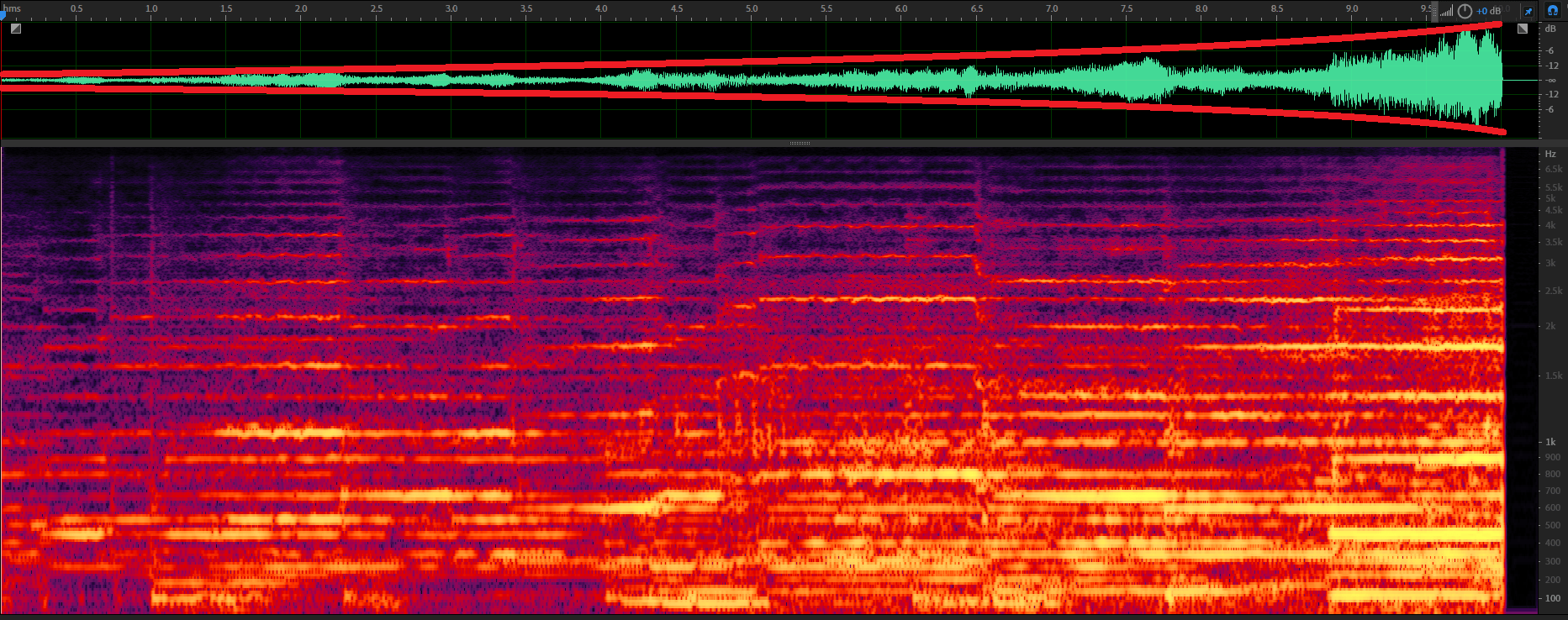}
    \caption{Mel-spectrogram of a classical music piece generated by \model{}. The effect of gradual volume increase (crescendo) is apparent (red color envelope around waveform).}
    \label{fig:sample_mustango3}
\end{figure}

\begin{figure}[h]
    \centering
    \includegraphics[width=0.7\textwidth]{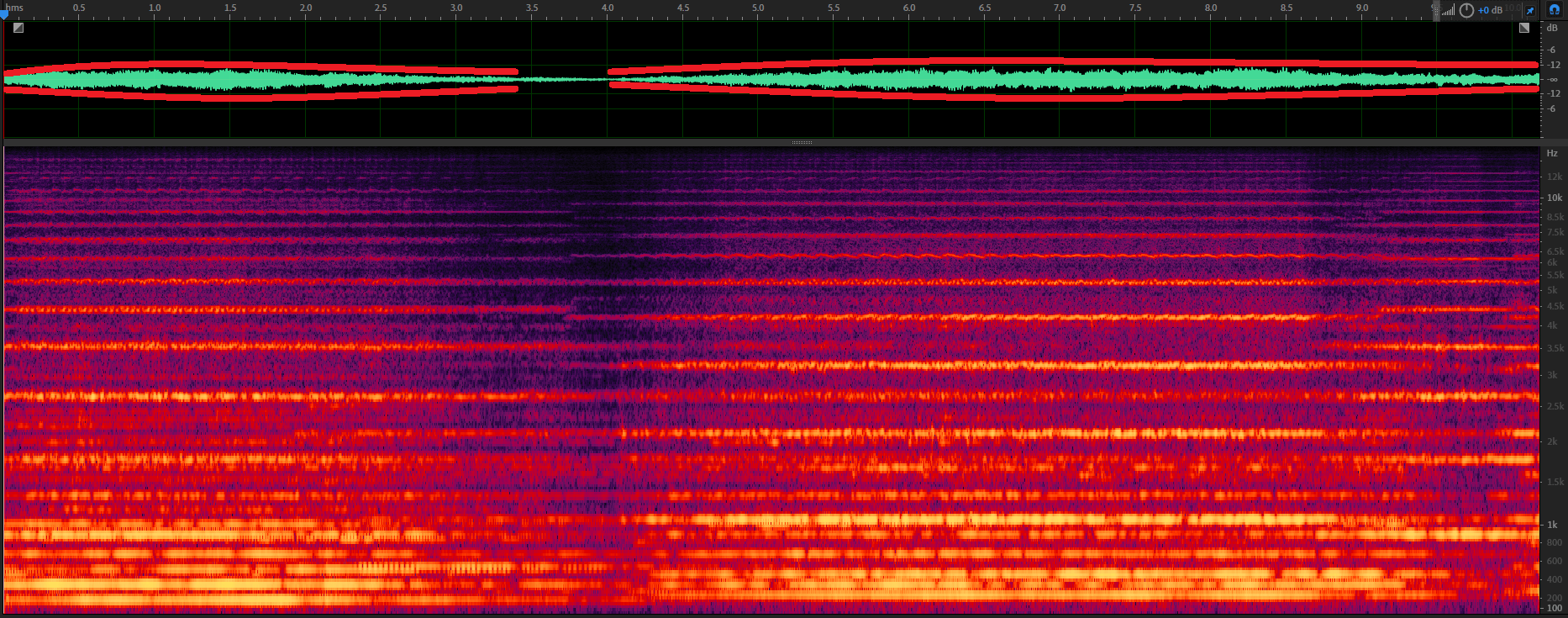}
    \caption{Mel-spectrogram of a classical music piece generated by \texttt{MusicGen-M}. The effect of gradual volume increase (crescendo) is not clear (red color envelope around waveform).}
    \label{fig:sample_musicgen3}
\end{figure}

\begin{figure}[h]
    \centering
    \includegraphics[width=0.7\textwidth]{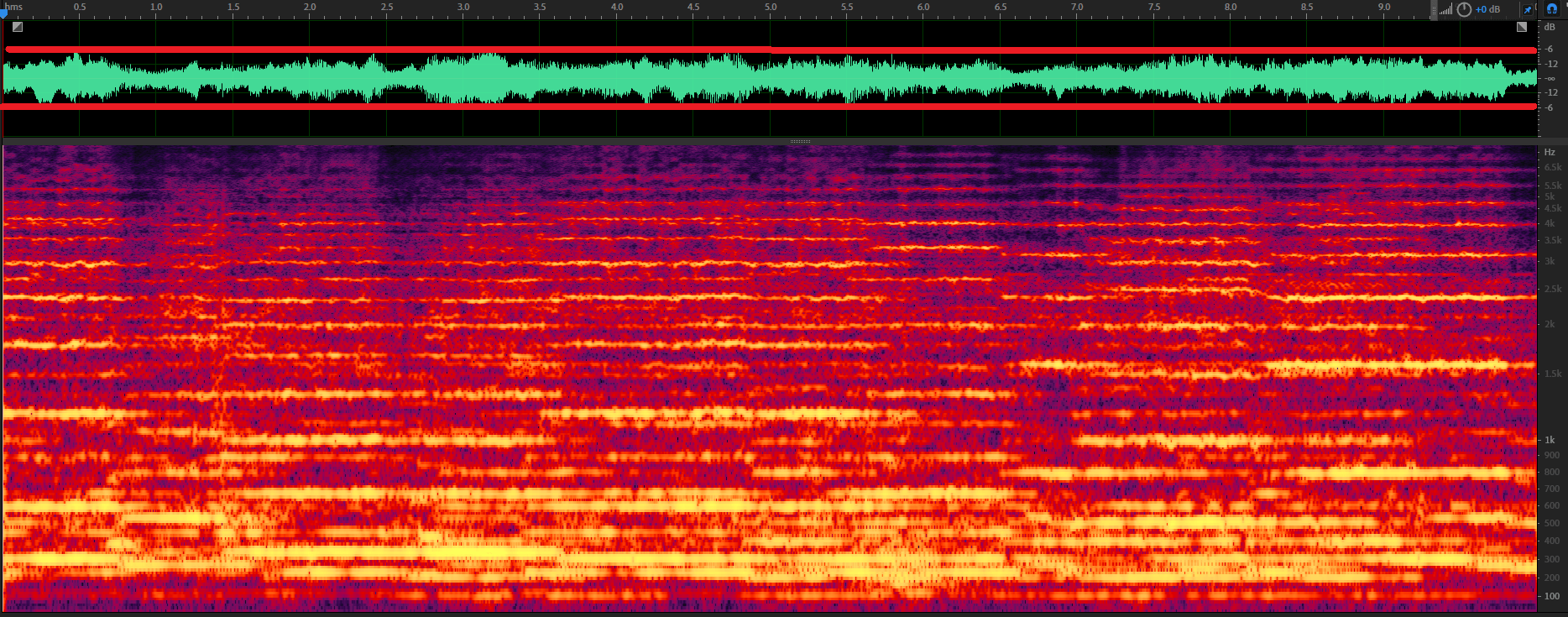}
    \caption{Mel-spectrogram of a classical music piece generated by \texttt{AudioLDM-2}. The effect of gradual volume increase (crescendo) is not present (red color envelope around waveform).}
    \label{fig:sample_audioldm3}
\end{figure}







\end{document}